\documentclass[a4paper, oneside, fleqn, 12pt]{scrartcl}

\usepackage[ngerman,USenglish]{babel}
\usepackage[utf8]{inputenc}
\usepackage[authoryear,round,colon]{natbib}
\usepackage{amsmath,amssymb,bm}
\usepackage{textcomp}
\usepackage{dsfont}
\usepackage{color}
\usepackage{graphicx}
\usepackage{graphics}
\usepackage{multirow}
\usepackage{lscape}
\usepackage{tabularx}
\usepackage{tabulary}
\usepackage{booktabs}
\usepackage{authblk}
\usepackage{subfigure}
\usepackage[font={small,sl},labelfont=bf]{caption} 

\usepackage[running]{lineno}
\usepackage[pagebackref]{hyperref}
\hypersetup{%
	colorlinks=true,
	linkcolor=blue, 
	citecolor=blue,
	bookmarksopen=true,
	pdfstartpage={1},
	pdftitle={Estimating timber volume loss due to storm damage in Carinthia, Austria, using ALS/TLS and spatial regression models},
	pdfauthor={Arne Nothdurft, Christoph Gollob, Ralf Kraßnitzer, Gernot Erber, Tim Ritter, Karl Stampfer, Andrew O. Finley},
	pdfkeywords={}
}

\newcommand{\bI}{ {\boldsymbol I} }

\newcommand{\bs}{ {\boldsymbol s} }

\newcommand{\bw}{ {\boldsymbol w} }

\newcommand{\bX}{ {\boldsymbol X} }
\newcommand{\by}{ {\boldsymbol y} }

\newcommand{\bphi}{ {\boldsymbol \phi} }

\newcommand{\bbeta}{ {\boldsymbol \beta} }
\newcommand{\btheta}{ {\boldsymbol \theta} }
\newcommand{\bSigma}{ {\boldsymbol \Sigma} }

\newcommand{\calA}{\mathcal{A}}

\newcommand{\calB}{\mathcal{B}}

\newcolumntype{L}[1]{>{\raggedright\arraybackslash}p{#1}} 
\newcolumntype{C}[1]{>{\centering\arraybackslash}p{#1}} 
\newcolumntype{R}[1]{>{\raggedleft\arraybackslash}p{#1}} 

\newcommand{\beginsupplement}{%
        \setcounter{table}{0}
        \renewcommand{\thetable}{S\arabic{table}}%
        \setcounter{figure}{0}
        \renewcommand{\thefigure}{S\arabic{figure}}%
}

\graphicspath{{Graph/}}

\title{\Large 
Estimating timber volume loss due to storm damage in Carinthia, Austria, using ALS/TLS and spatial regression models
}

\author[1]{\large Arne Nothdurft\thanks{\href{mailto:arne.nothdurft@boku.ac.at}{arne.nothdurft@boku.ac.at}, Tel: +43-1-47654-91411}}
\author[1]{\large Christoph Gollob}
\author[1]{\large Ralf Kraßnitzer} 
\author[2]{\large Gernot Erber} 
\author[1]{\large Tim Ritter} 
\author[2]{\large Karl Stampfer} 
\author[3]{\large Andrew O. Finley}

\affil[1]{\normalsize University of Natural Resources and Life Sciences, Vienna (BOKU), Department of Forest and Soil Sciences, Institute of Forest Growth, Austria}

\affil[2]{\normalsize University of Natural Resources and Life Sciences, Vienna (BOKU), Department of Forest and Soil Sciences, Institute of Forest Engineering, Austria}

\affil[3]{\normalsize Department of Forestry, Michigan State University, East Lansing, MI 48824-1222, USA}

\date{\large \today}

\begin{document}

\maketitle


\begin{abstract}
A spatial regression model framework is presented to predict growing stock volume loss due to storm Adrian which caused heavy forest damage in the upper Gail valley in Carinthia, Austria, in October 2018. Model parameters were estimated using growing stock volume measured with a terrestrial laser scanner on 62 sample plots distributed across five sub-regions. Predictor variables were derived from high resolution vegetation height measurements collected during an airborne laser scanning campaign. Non-spatial and spatial candidate models were proposed and assessed based on fit to observed data and out-of-sample prediction. Spatial Gaussian processes associated model intercepts and regression coefficients were used to capture spatial dependence. Results show a spatially-varying coefficient model, which allowed the intercept and regression coefficients to vary spatially, yielded the best fit and prediction. Two approaches were considered for prediction over blowdown areas: 1) an \emph{areal} approach that viewed each blowdown as a single prediction unit indexed by its centroid; and 2) a \emph{block} approach where each blowdown was partitioned into smaller prediction units to better align with sample plots' spatial support. Joint prediction was used to acknowledge spatial dependence among block units. Results demonstrated the block approach is preferable as it mitigated change-of-support issues encountered in the areal approach. Despite the small sample size, predictions for 55\,\% of the total 564 blowdown areas, accounting for 93\,\% of the total loss, had a coefficient of variation less than 25\,\%. Key advantages of the proposed regression framework and chosen Bayesian inferential paradigm, were the ability to quantify uncertainty in spatial covariance parameters, propagate parameter uncertainty through to prediction, and provide statistically valid prediction point and interval estimates for individual blowdowns and collections of blowdowns at the sub-region and region scale via posterior predictive distribution summaries.

\vspace{0.25cm}

\end{abstract}

\section{Introduction}\label{sec:intro}

Analyses by \cite{Schelhaas2003} showed that storms were responsible for more than half the total damages in European forests for the period 1950--2000. The majority of this forest damage occurred in the Alpine zone of mountainous regions. In Austria, for the period 2002--2010, storms damaged 3.1 million m\textsuperscript{3} annually \citep{Thom2013}, representing 0.26\,\% of the total growing stock as well as 12\,\% of the total annual fellings. 
In Switzerland, the storm damage was 17 and 22 times higher in the period 1985--2007 than in the two preceding 50 years periods \citep{Usbeck2010}. 

According to \cite{Usbeck2010}, the possible explanations for such an increasing trend were manifold and include increased growing stocks, enlarged forested area, milder winters and hence tendency for wet and unfrozen soils, and higher recorded maximum gust wind speeds.
Since 1990, 85\,\% of all primary damage in the Western, Central and Northern European regions was caused by catastrophic storms with maximum gust wind speed between 50--60 ms\textsuperscript{-1} \citep{Gregow2017}. The relevance of storm damage is likely to increase, as simulations with climate scenario data and forest stand projections suggested a higher probability of exceeding the critical wind speed and, hence, wind throw events \citep{Blennow2008,Blennow2010}. Such a trend could negatively impact the carbon balance, especially in Western and Central European regions \citep{Lindroth2009}.

Wind throw events have direct and indirect impact on long-term sustained yield planning. A common indirect cost stems from the fact that storm-felled trees provided a surplus of breeding material for bark beetles and promoted their rapid population increase causing extra timber losses in the subsequent years \citep{Marini2017}. Hence, it is recommended that fallen trees be removed within two years of the disturbance \citep{DeGroot2018}. Following a wind throw event and prior to harvesting the storm-felled trees, strategic salvage harvest planning is needed \citep{Jirikowski2003}. Implementing such plans requires unplanned/unbudgeted logging road and site development efforts as well as interaction with external harvesting, transportation, and processing industry. 

To inform strategic salvage harvest planning, a rapid and accurate estimate of the spatial extent and local severity of damage is needed. The task is to provide such estimates for individual and collections of discrete blowdowns. In most situations, a paucity of existing field inventory data within or adjacent to blowdowns precludes design-based estimation. Rather, field data from an existing set of sparsely sampled inventory plots or a small purposively selected set of sample plots, can be coupled with auxiliary data using a model to yield viable estimates. In such settings, the forest response variable measured on sample plots is regressed against meaningful auxiliary data. Commonly, such auxiliary data come as remotely sensed variables collected via satellite, aircraft, or unmanned aerial vehicle (UAV) based sensors. Remotely sensed data are routinely used to support forest inventories and have been shown to increase accuracy and precision of the estimates, and reduce field data collection efforts; see \cite{Koehl2006} and other references herein.

Increasingly, laser imaging detection and ranging (LiDAR) data are being incorporated into forest inventory and mapping efforts. LiDAR measurements offer high-resolution and 3-dimensional (3D) representation of forest canopy structure metrics that are often related to forest variables of interest. For example, LiDAR height metrics have been successfully used to estimate average tree height \citep{MagnussenBoudewyn1998}, stem density \citep{NaessetBjerknes2001}, basal area \citep{Naesset2002,Magnussen2010}, biomass \citep{Finley2017, Babcock2018} and growing stock volume \citep{Nelson1988,Naesset1997,Maltamo2006a}.

In practice, different techniques are used to predict forest variables using field measurements and remote sensing data. Nonparametric imputation methods, using some flavor of nearest neighbor (NN) interpolation \citep{MoeurStage1995}, achieved robust prediction for possibly multivariate response variable vectors \citep{TomppoHalme2004,LeMayTemesgen2005,Maltamo2006b,PackalenMaltamo2007}. \cite{Hudak2008} demonstrated the NN prediction accuracy can be enhanced through resampling and classifications with ``random forests'' \citep{Breiman2001}.
A key shortcoming of NN imputation methods, however, is the lack of statistically robust variance estimates, with the exception of some approximations presented and assessed in \cite{McRoberts2007,McRoberts2011} and \cite{Magnussen2013}.

Similar to NN imputation methods, geostatistical methods yield accurate and precise spatial prediction of forest variables. Additionally, geostatistical methods provide a solid theoretical foundation for probability-based uncertainty quantification, see, e.g., \cite{VerHoef2013}. In such settings, point-referenced observations are indexed by spatial locations, e.g., latitude and longitude, and predictive models build on classical kriging methods \citep{Cressie1993,Schabenberger2004,Chiles2013}. Additional flexibility for specifying and estimating regression mean and covariance components comes from recasting these predictive models within a Bayesian inferential framework \citep{Banerjee2014}. Similar extensions and benefits have been demonstrated for data with observations indexed by areal units, particularly in small-area estimation settings, many of which build on the Fay-Harriot model \citep{FayHerriot1979}, see, e.g., \cite{VerPlanck2017, VerPlanck2018} and references therein. 

Following the study region description and data in Section~\ref{sec:data}, a general model that subsumes various sub-models is developed in Section~\ref{sec:model} along with implementation details for parameter estimation, prediction, and model selection in Section~\ref{sec:implementation}. Here too, we offer two different approaches for prediction over the areal blowdown units. Results presented in Section~\ref{sec:results} focus first on model selection then comparison among blowdown predictions, which is followed by discussion in Section~\ref{sec:discussion}. Some final remarks and future direction are provided in Section~\ref{sec:conclusion}.

\section{Materials and methods}\label{sec:methods}

\subsection{Study region and model data}\label{sec:data}

\begin{figure}[!ht]
\begin{center}
	\subfigure[]{\includegraphics[width=10cm,trim={0cm 0cm 0cm 0cm},clip]{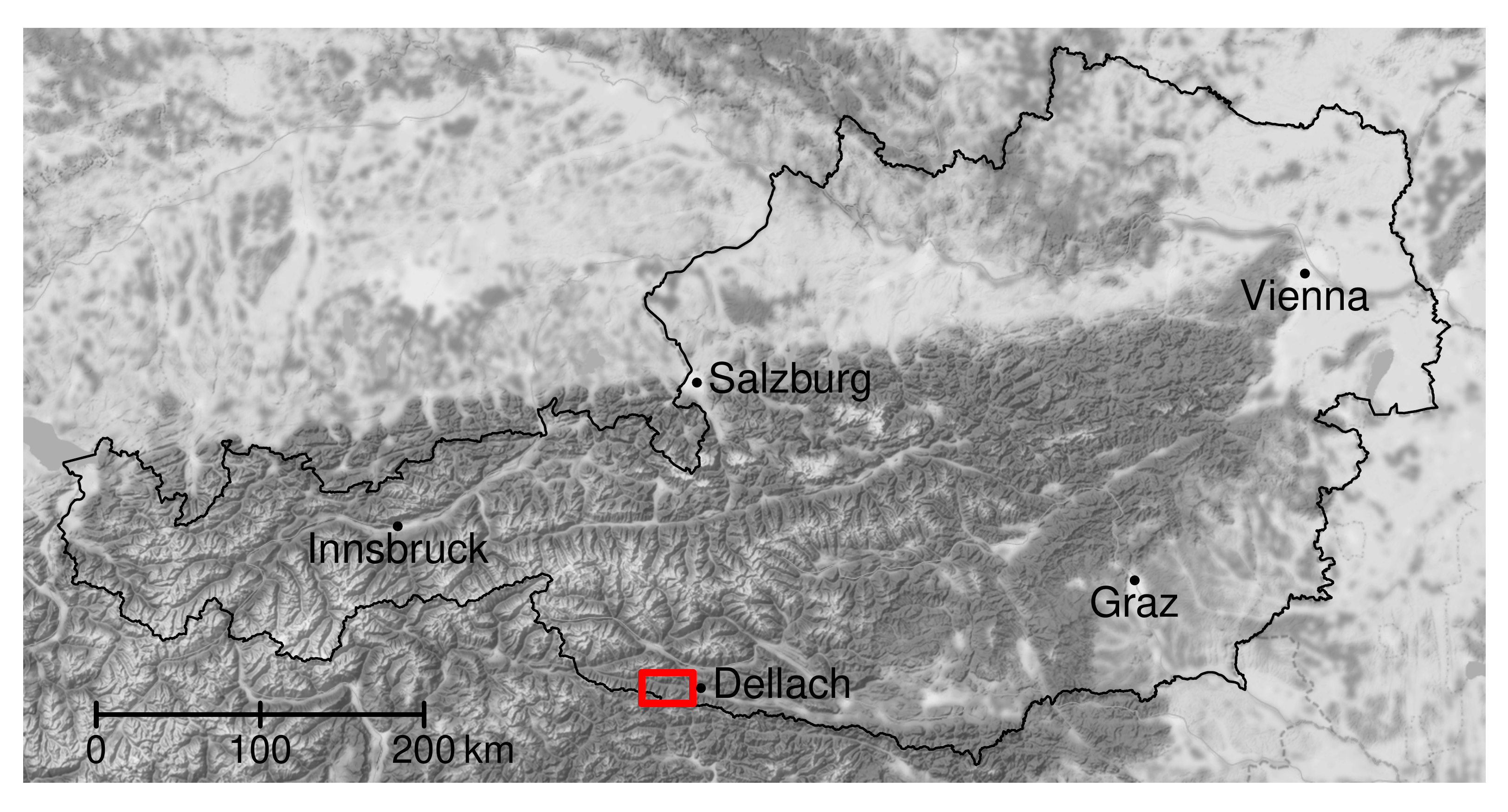}\label{fig:austria}}\\
	\subfigure[]{\includegraphics[width=10cm,trim={0cm 0cm 0cm 0cm},clip]{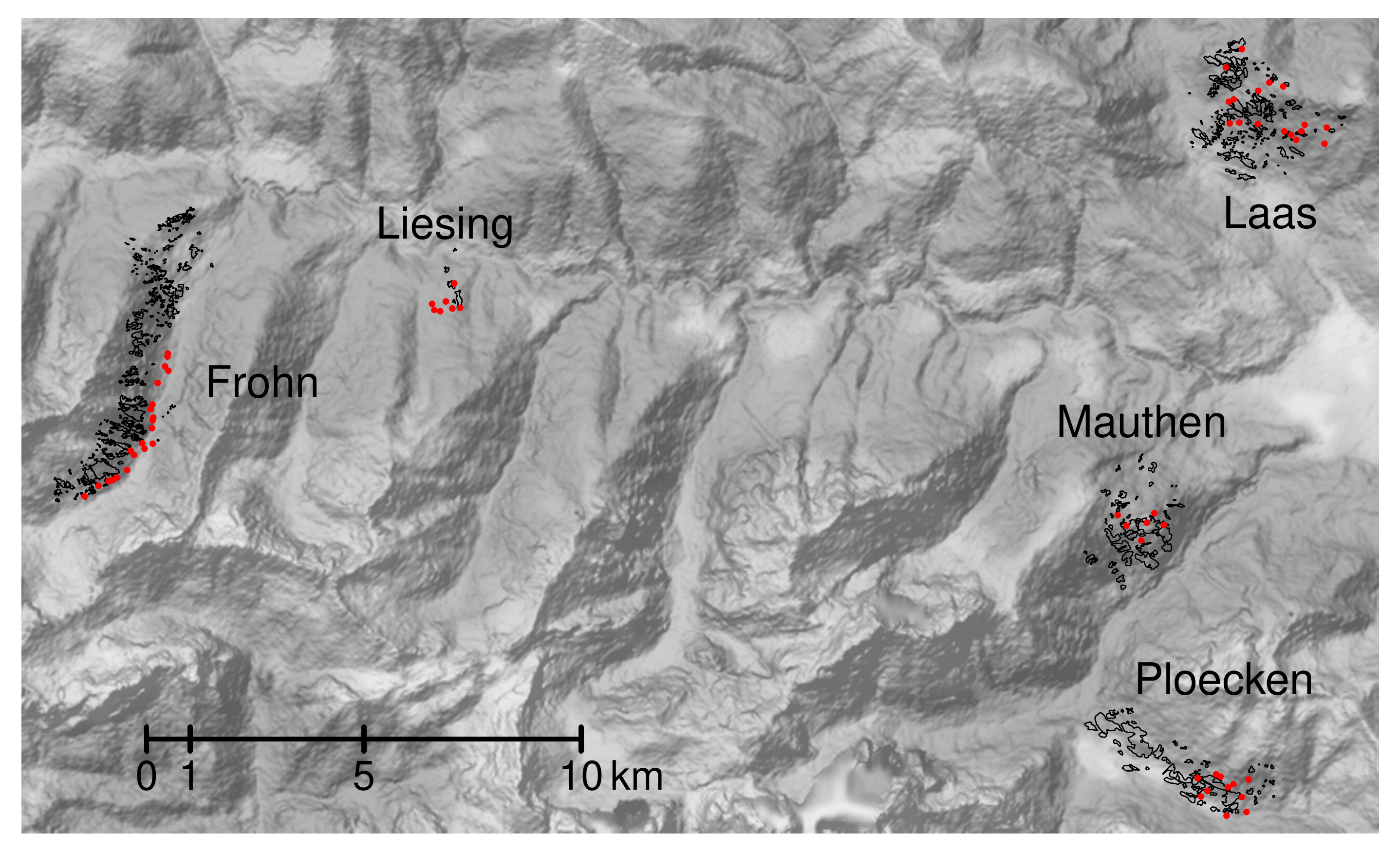}\label{fig:samplePlots}}
	\caption{\subref{fig:austria} Location of the study region in Southern Carinthia, Austria. \subref{fig:samplePlots} Storm damage areas (polygons) and sample plots (red dots) in the study region.} \label{fig:StudySite1}
\end{center}
\end{figure}

The study region was located in the southern region of the Austrian federal state of Carinthia, within the upper Gail valley and near the Dellach forest research and training center, which is jointly operated by the Institute of Forest Growth and the Institute of Forest Engineering of the University of Natural Resources and Life Sciences Vienna (Figure~\ref{fig:austria}). On October 28, 2018, the storm Adrian formed over the western Mediterranean Sea and achieved wind gust speeds of 130\,km/h throughout Carinthia. Within 72 hours, the storm was producing 627\,\,l/m$^2$ of precipitation at the Pl{\"o}ckenpass meteorological station, located near the study region, and inflicting heavy damage on the region's forest \citep{Zimmermann2018}. 

Forest blowdown, caused by Adrian, was widely distributed across the Hermagor administrative district (Bezirk), which covers the study region. Using high-resolution aerial images provided by the Carinthian Forest Service, blowdown occurrences in the district were identified and delineated with high accuracy. As illustrated in Figure~\ref{fig:samplePlots} the blowdowns were concentrated in five distinct sub-regions labeled Frohn, Laas, Liesing, Mauthen, and Ploecken. A total of 564 blowdowns were delineated, totalling 212.3\,ha of affected forest. Table~\ref{tab:SumStatAreaPlot} provides the number of blowdown occurrences across the sub-regions and affected area characteristics.

\begin{table}
\caption{Summary statistics of the digitized blowdowns and collected sample plot data by sub-region.}\label{tab:SumStatAreaPlot}
\small
\setlength{\tabcolsep}{0.5pt}
\begin{tabularx}{1.0\linewidth}{
L{0.075\linewidth}
R{0.070\linewidth}
R{0.075\linewidth}
R{0.075\linewidth}
R{0.075\linewidth}
R{0.075\linewidth}
R{0.075\linewidth}
R{0.070\linewidth}
R{0.075\linewidth}
R{0.075\linewidth}
R{0.075\linewidth}
R{0.075\linewidth}
R{0.075\linewidth}
}
& \multicolumn{6}{c}{Blowdown areas} & \multicolumn{6}{c}{Sample plots}\\
\cmidrule(lr){2-7} \cmidrule(lr){8-13}
& & \multicolumn{5}{c}{Area (ha)} & & \multicolumn{5}{c}{Growing stock volume (m$^3$/ha)}\\
\cmidrule(lr){3-7} \cmidrule(lr){9-13}
 & n & mean & med & sd & min & max & N & mean & med & sd & min & max \\ 
\midrule
Frohn & 273 & 0.230 & 0.065 & 0.600 & 0.004 & 7.231 & 21 & 593 & 579 & 227 & 188 & 979 \\ 
Laas & 152 & 0.362 & 0.138 & 0.565 & 0.004 & 2.727 & 17 & 785 & 772 & 333 & 142 & 1382 \\ 
Liesing & 5 & 0.581 & 0.348 & 0.621 & 0.115 & 1.607 & 7 & 689 & 698 & 67 & 563 & 757 \\ 
Mauthen & 62 & 0.621 & 0.190 & 1.075 & 0.013 & 4.969 & 6 & 762 & 748 & 323 & 403 & 1220 \\ 
Ploecken & 72 & 0.738 & 0.242 & 1.724 & 0.008 & 12.322 & 11 & 773 & 773 & 136 & 528 & 968 \\ 
\midrule
Total & 564 & 0.376 & 0.110 & 0.892 & 0.004 & 12.322 & 62 & 705 & 730 & 256 & 142 & 1382 \\  
\end{tabularx}
\end{table}

Forest inventory data were not available for the study region; therefore, a field measurement campaign was initiated in May 2020 (post-Adrian) to collect growing stock timber volume measurements suitable for estimating volume loss due to blowdown. A total of $n$=62 sample plots were installed in unaffected forest adjacent to blowdowns (Figure~\ref{fig:samplePlots}). Plot locations were chosen to characterize forest similar in structure and composition to the pre-blowdown forest. No plots were located within blowdowns. Plot measurements were conducted using the terrestrial laser scanning (TLS) GeoSLAM ZEB HORIZON system (GeoSLAM Ltd., Nottingham, UK). Position and diameter at breast height (DBH) for the approximately 5586 measurement trees were derived from 3D point clouds on a 20\,m radius plot using fully automated routines demonstrated in \cite{Gollob2019,Gollob2020} and in \cite{Ritter2017,Ritter2020}. Tree height was estimated using a N\"{a}slund function formulated as a mixed-effects model with plot-level random effects. Stem volume was calculated using a traditional stem-form function \citep{Pollanschuetz1965}. For each plot, growing stock timber volume was expressed as m$^3$/ha (i.e., computed as the sum of tree volume scaled by the 7.958 fixed-area plot tree expansion factor). Table~\ref{tab:SumStatAreaPlot} provides a summary of growing stock timber volume by sub-region, the values of which serve as the response variable observations in  subsequent regression models.

The Carinthian Forest Service provided a comprehensive set of aerial laser scanning (ALS) variables summarized on a 1\,m $\times$ 1\,m resolution grid collected in 2012 over the study region. Within each grid cell, ALS variables comprised the point cloud height distribution's mean, median, min, maximum, and standard deviation. Values for each variable in this fine grid were averaged over each plot to yield a set of plot-level predictor variables to pair with the growing stock volume response variable.

\subsection{Model construction} \label{sec:model}

As stated in Section~\ref{sec:intro}, the study goal was to predict what 2020 growing stock timber volume in blowdowns would have been if not destroyed by Adrian in late October 2018. Sample plot data, which included 2020 response variable measurements and 2012 ALS predictor variables, were used to develop models to predict growing stock timber volume for blowdowns where only 2012 ALS measurements were available. 

Due to the relatively small number of sample plots within any one of the five spatially disjoint sub-regions (Figure~\ref{fig:StudySite1}), we aimed to pool the 62 sample plot measurements. An ideal pooled model would allow for intra- and inter-location specific relationships between the response and predictor variables. Such spatially varying relationships are particularly attractive because they accommodate potential impact of unobserved (and for the most part unobservable) spatially-explicit mediating factors such as disturbance history, genetics, and local growth environments. When cast within a regression framework, a spatially varying coefficients (SVC) model uses a smoothly-varying spatial process to pool information and avoid overfitting \citep{Gelfand2003, Finley2011}. 

Given prediction is our primary goal, a preferred model would also estimate residual spatial correlation (i.e., spatial dependence among measurements not explained by the predictor variables) and use it to improve prediction performance. We anticipated the residual spatial correlation would decrease as distance between measurements increased. Again, following \cite{Gelfand2003} and broader geostatistical literature cited in Section~\ref{sec:intro}, residual spatial dependence is effectively estimated using a spatial process. 

To accommodate the anticipated data features, assess varying levels of model complexity, and deliver statistically valid probabilistic uncertainty quantification we model response $y(\bs)$ at generic spatial location $\bs$ as
\begin{linenomath*} 
\begin{equation}\label{eq: spatially_varying_regression}
 y(\bs) = (\beta_0 + \delta_0w_0(\bs)) + \sum_{j=1}^{p} x_j(\bs)\left\{\beta_j + \delta_jw_j(\bs)\right\} + \epsilon(\bs),
\end{equation}
\end{linenomath*} 
where $x_j(\bs)$, for each $j=1,\ldots,p$, is the known value of a predictor variable at location $\bs$, $\beta_j$ is the regression coefficient corresponding to $x_j(\bs)$, $\beta_0$ is an intercept, and $\epsilon(\bs)$ follows a normal distribution with mean zero and variance $\tau^2$. Here, $\tau^2$ is viewed as the measurement error variance. The quantities $w_0(\bs)$ and $w_j(\bs)$ are spatial random effects corresponding to the intercept and predictor variables, respectively, thereby yielding a spatially varying regression model. To allow for varying levels of model complexity, the  $\delta$s in Equation~(\ref{eq: spatially_varying_regression}) are binary indicator variables used to turn on and off the spatial random effects (i.e., a value of $1$ means the given spatial random effect is included in the model and $0$ otherwise). When $\delta = 1$ the associated space-varying regression coefficient can be seen as having a global effect, i.e., $\beta_0$ or $\beta_j$s, with local adjustments, i.e., $w_0(\bs)$ or $w_j(\bs)s$.

Over $n$ locations, a given spatial random effect $\bw = (w(\bs_1), w(\bs_2), w(\bs_3), \ldots, w(\bs_n))^\top$ follows a multivariate normal distribution with a length $n$ zero mean vector and $n\times n$ covariance matrix $\bSigma$ with $(i,j)^{th}$ element given by $C(\bs_i, \bs_j; \btheta)$. Clearly, for any two generic locations $\bs$ and $\bs^\ast$ locations within the study region the function used for $C(\bs, \bs^\ast; \btheta)$ must result in a symmetric and positive definite matrix $\bSigma$. Such functions are known as positive definite functions, details of which can be found in  \cite{Cressie1993}, \cite{Chiles2013}, and \cite{Banerjee2014}, among others. Here we specify $C(\bs, \bs^\ast; \btheta) = \sigma^2\rho(\bs, \bs^\ast; \bphi)$ where $\btheta = \{\sigma^2, \bphi\}$ and $\rho(\cdot ; \bphi)$ is a positive support correlation function with $\bphi$ comprising one or more parameters that control the rate of correlation decay and smoothness of the process. The spatial process variance is given by $\sigma^2$, \textit{i.e.}, $\text{Var}(w(\bs))=\sigma^2$. This covariance function yields a \emph{stationary} and \emph{isotropic} process, \textit{i.e.}, a process with a constant variance and a correlation depending only on the Euclidean distance separating locations. The M{\'a}tern correlation function is a flexible class of correlation functions with desirable theoretical properties \citep{Stein1999} and is given by
\begin{linenomath*} 
\begin{equation}\label{matern}
    \rho(||\bs-\bs^\ast||; \bphi) = \frac{1}{2^{\nu - 1}\Gamma(\nu)}(\phi||\bs - \bs^\ast||)^\nu \mathcal{K}_\nu(||\bs - \bs^{\ast}||; \phi);\; \phi > 0, \; \nu > 0,
\end{equation}
\end{linenomath*} 
where $||\bs-\bs^\ast||$ is the Euclidean distance between $\bs$ and $\bs^\ast$, $\bphi = \{\phi, \nu\}$ with $\phi$ controlling the rate of correlation decay and $\nu$ controlling the process smoothness, $\Gamma$ is the Gamma function, and $\mathcal{K}_\nu$ is a modified Bessel function of the third kind with order $\nu$. While it is theoretically ideal to estimate both $\phi$ and $\nu$, it is often useful from a computational standpoint to fix $\nu$ and estimate only $\phi$. For our current analysis, such a concession is reasonable given there is likely little information gain in estimating both parameters. Conveniently, when $\nu=0.5$ the M{\'a}tern correlation reduces to the exponential correlation function, i.e., $\rho(||\bs-\bs^\ast||; \phi) = \exp(-\phi ||\bs-\bs^\ast||)$. Therefore, only two process parameters are estimated for any given random effect are $\btheta=\{\sigma^2, \phi\}$.

In the subsequent analysis, we consider the following candidate models defined using the general model (\ref{eq: spatially_varying_regression}):
\begin{enumerate}
    \item Non-spatial---all $\delta$ are set to zero. This is simply a multiple regression model.
    \item Space-varying intercept (SVI)---$\delta_0$ equals $1$ and all other $\delta$s are set to zero. This model estimates a spatial process and associated parameters $\btheta_0$ for the intercept, but the impact of the covariates is assumed to be the same across the study region.
    \item Space-varying coefficients (SVC)---all $\delta$s are set to 1. This model allows all regression coefficients to vary spatially over the study region. Each spatial process has its own parameters $\btheta_0$ and $\btheta_j$ for $j = (1,2, \ldots, p)$.
\end{enumerate}

\subsection{Implementation and analysis}\label{sec:implementation}

To facilitate uncertainty quantification for model parameters and subsequent prediction, the Non-spatial, SVI, and SVC candidate models defined in Section~\ref{sec:model} were fit within a Bayesian inferential framework, see, e.g., \cite{Gelman2013} for a general description Bayesian model fitting methods. The candidate models' Bayesian specification is completed by assigning prior distributions to all parameters. Then, parameter inference follows from posterior distributions that are sampled using Markov chain Monte Carlo (MCMC) algorithms. \cite{Finley2020} provide open source software, available in \texttt{R} \citep{R}, that implement efficient MCMC sampling algorithms for Equation~(\ref{eq: spatially_varying_regression}) and associated sub-models. More specifically, the \texttt{spSVC} function within the \texttt{spBayes} package \citep{spBayes} provides MCMC-based parameter posterior summaries, fit diagnostics for model comparison, and prediction at unobserved locations. 

\emph{Data and annotated \texttt{R} code needed to fully reproduce the analysis and results will be provided as Supplementary Material upon publication or if requested by reviewers.}

\subsubsection{Prediction}\label{sec:prediction}

Our interest is in predicting the 2020 growing stock timber volume response variable $\tilde{\by} = (\tilde{y}(\bs_1), \tilde{y}(\bs_2), \ldots, \tilde{y}(\bs_{n_0}))^\top$ at a set of $n_0$ locations where it is not observed but ALS predictors are available (we used tilde to indicate a prediction). Following \cite{Gelman2013} and \cite{Banerjee2014}, given MCMC samples from the posterior distributions of the posited model's parameters, composition sampling is used to sample one-for-one from $\tilde{\by}$'s posterior predictive distribution (PPD). For example, the Non-spatial model's $l$-th PPD sample $\tilde{\by}^{(l)}$ is drawn from the multivariate Normal distribution $MVN\left(\beta_0^{(l)} + \bX\bbeta^{(l)}, \tau^{2(l)}\bI\right)$, where $\beta_0^{(l)}$, $\bbeta^{(l)} = \left(\beta_1^{(l)}, \beta_2^{(l)}, \ldots, \beta_p^{(l)}\right)^\top$, and $\tau^{2(l)}$ are the $l$-th joint samples from the parameters' posterior distribution, $\bX$ is the $n_0\times p$ matrix of predictors at the $n_0$ prediction locations, and $\bI$ is the $n_0\times n_0$ identity matrix. The multivariate Normal PPDs for the SVI and SVC models are given in \cite{Finley2020, spBayes}. Importantly, the SVI and SVC candidate models use joint composition sampling to acknowledge the spatial correlation among prediction locations. A given candidate model's PPD is evaluated using each of its parameter's $M$ posterior samples, i.e., $l = 1, 2, \ldots, M$, to generate $M$ PPD samples for each of the $n_0$ prediction locations. These PPD samples are summarized analogously to the parameters' posterior samples. Prediction point estimates could include the PPD mean or median, and interval estimates can be built off the PPD standard deviation or directly expressed using a set of lower and upper percentiles in the form of a credible interval. For example, a point and dispersion estimate for the growing stock timber volume at a prediction location could be the mean and standard deviation over that location's $M$ PPD samples.

\subsubsection{Model selection}\label{sec:selection}

Our analysis had two separate model selection steps. First, in an effort to build parsimonious models and reduce possible issues arising from collinearity among the often highly correlated ALS predictor variables, we identified a common set of predictors to use in the three candidate models. Given our focus on prediction, predictor variable selection aimed to minimize prediction error using leave-one-out (LOO) cross-validation among the 62 sample plot measurements. Second, given the common set of predictors, model fit and prediction performance were used to select the ``best'' candidate model for subsequent blowdown area prediction. Here, again, candidate model prediction performance was assessed using LOO cross-validation among the 62 sample plot measurements. 

For the first model selection step, the Non-spatial model was used to select the set of ALS predictor variables that minimized LOO mean squared prediction error (MSPE). As described in Section~\ref{sec:data}, candidate variables were summaries of the ALS point cloud height distribution and included its mean, median, minimum, maximum, and standard deviation. A backward variable selection algorithm, implemented using the \texttt{leaps} \citep{leaps} and \texttt{caret} \citep{caret} \texttt{R} packages, identified the predictor set that minimized LOO MSPE. 

The second model selection step used model fit and LOO prediction measures to identify the ``best'' model for subsequent blowdown area prediction. The deviance information criterion (DIC)\citep{spiegelhalter2002} and widely applicable information criterion (WAIC) \citep{Watanabe2010} model fit criterion were computed for each candidate model. DIC equals $-2(\text{L}-\text{p}_D)$ where L is goodness of fit and p$_D$ is a model penalty term viewed as the effective number of parameters. Two WAIC criteria were computed based on the log pointwise predictive density (LPPD) with $\text{WAIC}_1 = -2(\text{LLPD} - \text{p}_1)$ and $\text{WAIC}_2 = -2(\text{LLPD} - \text{p}_2)$ where penalty terms p$_1$ and p$_2$ are defined by \cite{Gelman2014} just prior to, and in, Equation~(\ref{eq: spatially_varying_regression}). Models with lower values of DIC, WAIC$_1$, and WAIC$_2$ have better fit to the observed data and should yield better out-of-sample prediction, see \cite{Gelman2014}, \cite{Vehtari2017}, or \cite{Green2020} for more details on these model fit criteria.

The three candidate models were also assessed based on LOO cross-validation prediction MSPE and continuous rank probability score (CRPS) \citep{Gneiting2007}. Because MSPE measures only predictive accuracy, CRPS is a preferable measure of predictive skill because it favors models with both high accuracy and precision. Models with lower MSPE and CRPS should yield better blowdown area predictions. In addition to MSPE and CRPS, we computed the percent of holdout observations covered by their corresponding PPD 95\% credible interval. Models with an empirical coverage percent close to the chosen PPD credible interval percent are favored.

\subsubsection{Blowdown prediction}\label{sec:blowdownPrediction}

Given the sample plot dataset and posited model identified using methods in Section~\ref{sec:selection}, two approaches were used to predict growing stock timber volume for blowdown areas. We refer to the approaches as \emph{areal} and \emph{block} prediction. 

\begin{figure}[!ht]
\begin{center}
	\subfigure[Areal mean canopy height.]{\includegraphics[width=7cm,trim={0cm 3.75cm 0cm 2.75cm},clip]{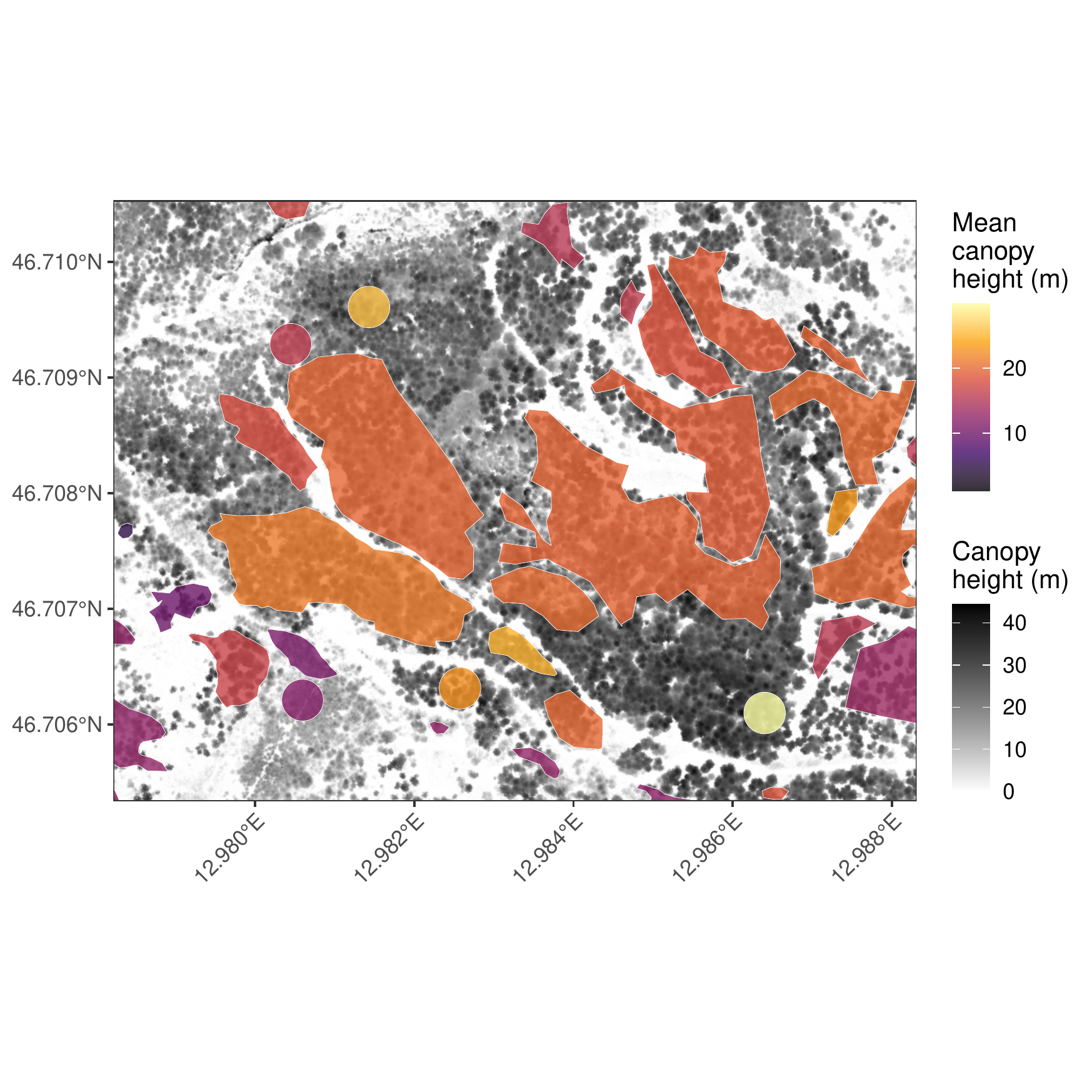}\label{fig:computeXAreal}}
	\subfigure[Grid cell mean canopy height.]{\includegraphics[width=7cm,trim={0cm 3.75cm 0cm 2.75cm},clip]{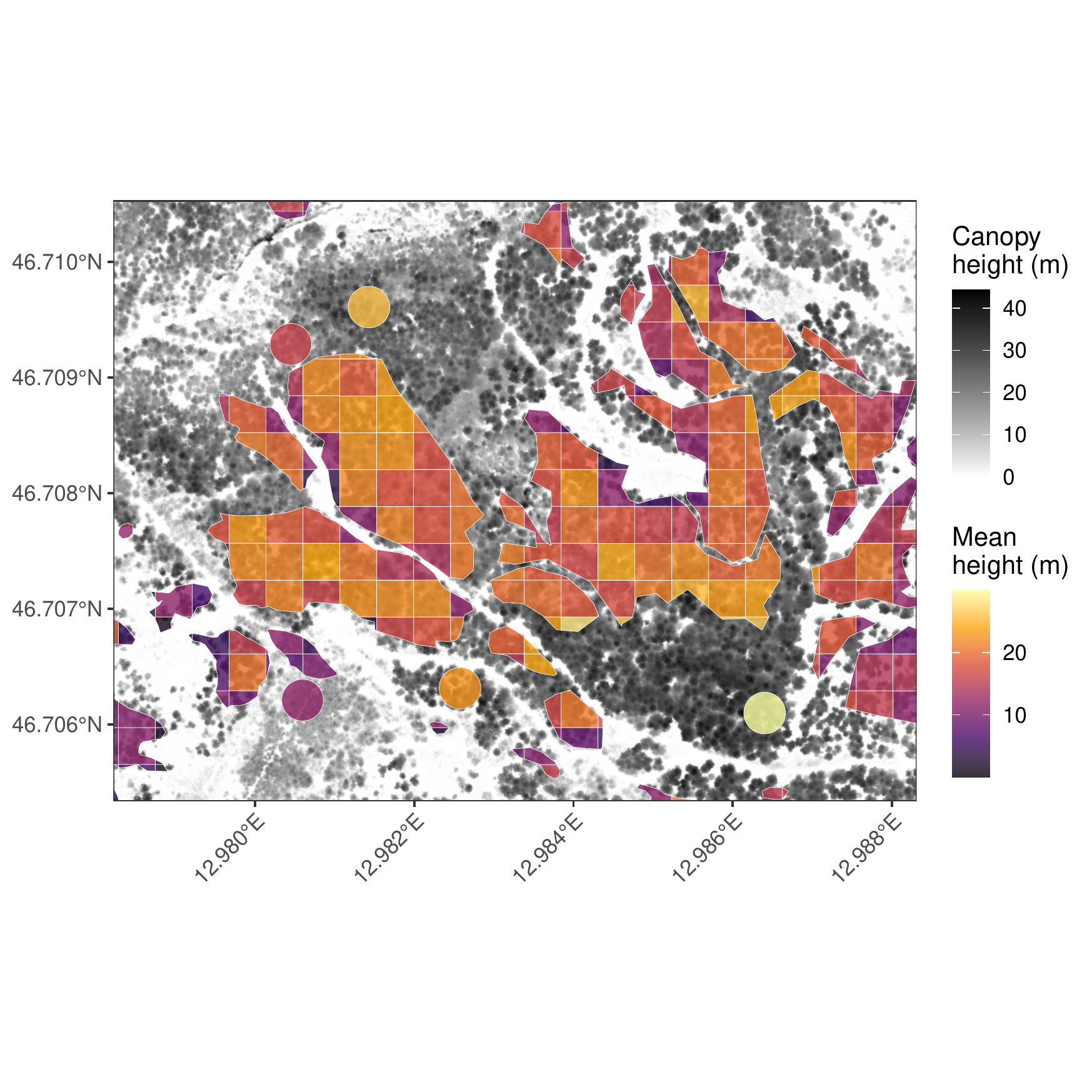}\label{fig:computeXBlock}}
	\caption{Illustration of the mean canopy height ALS predictor variable used for areal \subref{fig:computeXAreal} and \subref{fig:computeXBlock} block prediction for a subset of blowdowns in a small section of the Laas sub-region. The polygons are blowdowns and circles are forest sample plots. The gray scale basemap depicts the 1\,m $\times$ 1\,m ALS canopy height grid used to compute the mean canopy height values over plots and blowdown prediction units.} \label{fig:computeX}
\end{center}
\end{figure}

The \emph{areal} approach views each blowdown as a single prediction unit indexed by its polygon centroid and predictor variables computed as an average of the 1\,m $\times$ 1\,m resolution ALS values over its extent. Figure~\ref{fig:computeXAreal} illustrates the mean canopy height ALS variable summarized for blowdowns within a small portion of the Laas sub-region. \cite{VerPlanck2018} had a somewhat similar setting; however, their dataset and inferential goals differed from ours in a few key ways. First, because their sample data came from variable radius plots, it was not possible to spatially align ALS predictors with response variable measurements at the plot-level. As a result, they made the simplifying assumption that, when pooled, response variable measurements were representative of the stand areal unit within which they were observed. This assumption allowed them to spatially align the response and ALS predictor variables at the stand-level. Second, all stands within the population held at least two plots and the study goal was to improve stand-level point and interval estimates through a smoothing conditional autoregressive model (CAR) model. In our current setting, the sample data comprise response variable measurements collected on fixed-area plots with clearly defined spatial support over which the ALS predictor variables were also computed---the response and predictor variable measurements are spatially aligned at the plot-level. Also, no sample plots fall within the blowdown prediction units; hence, our inferential goal is squarely on out-of-sample prediction. Despite these differences we pursue the \emph{areal} prediction and clearly acknowledge the change-of-support issue between the discrete fixed-area plot data used to fit the model and different spatial support of the prediction units (see, e.g., \cite{Schabenberger2004} for a thorough description of spatial change-of-support problems). While not appropriate from a statistical standpoint, we do see this approach used in practice and therefore include it here for comparison with the \emph{block} approach that mitigates change-of-support issues by better aligning the spatial support of the measurement and prediction units.

For the areal approach, the $l$-th PPD sample of total growing stock volume (m$^3$) for a given blowdown is $\tilde{y}^{(l)}_{\calA} = A\tilde{y}(\bs)^{(l)}$, where $A$ is the blowdown's area (ha) and $\tilde{y}(\bs)^{(l)}$ is the growing stock volume (m$^3$/ha) predicted at the blowdown's centroid $\bs$. 

The \emph{block} approach partitions each blowdown into grid cells with the same area as the fixed-area sample plots (i.e., 0.126 ha). Each cell is indexed using its centroid, and predictor variables are computed as an average of the 1\,m $\times$ 1\,m resolution ALS values over its extent; see illustration in Figure~\ref{fig:computeXBlock}. Akin to block kriging \citep{Wackernagel2003}, the response variable prediction for a given blowdown is an aggregate of multiple point predictions within the blowdown extent. More specifically, given a blowdown divided into $n_0$ cells and corresponding vector holding the $l$-th joint PPD sample $\tilde{\by}^{(l)}$ (m$^3$/ha), the PPD sample of total growing stock volume (m$^3$) for the blowdown is $\tilde{y}^{(l)}_{\calB} = \sum^{n_0}_{i=1}a(\bs_i)\tilde{y}(\bs_i)^{(l)}$, where $a(\bs_i)$ is the area of the $i$-th cell that falls within the blowdown ($a_i$ is at most 0.126 ha when the cell's extent is completely within the blowdown). 

Composition sampling was again used to generate $M$ samples from $\tilde{y}_{\calA}$'s and $\tilde{y}_{\calB}$'s PPD for each blowdown. These PPD samples were summarized analogously to the parameters’ posterior samples to yield prediction point, dispersion, and interval estimates for the 564 blowdowns. In addition to blowdown specific PPD summaries, extra composition sampling was used to estimate the total volume PPD by sub-region and region. 

\section{Results}\label{sec:results}

\subsection{Candidate models}

Following Section~\ref{sec:selection}, the model that yielded minimum LOO cross-validation MSPE included only the ALS point cloud height distribution mean predictor variable. We refer to this predictor variable as \emph{mean canopy height} and set it as $x_1$ in Equation~(\ref{eq: spatially_varying_regression}) for all candidate models. 

\begin{table}[ht!]
  \caption{Parameter posterior distribution median and 95\% credible interval for candidate models.}\label{tab:estimates}
  \begin{center}
    \begin{tabular}{lccc}
    \toprule
    & Non-spatial & SVI & SVC\\
    \midrule
    $\beta_0$ & 159.8 (59.6, 264.1) & 172.9 (70.4, 277.5) & 176.3 (69.3, 277.8)\\
    $\beta_1$ & 33.7 (27.9, 39.8) & 33.2 (27.2, 39) & 33 (2.2, 63.9)\\
    $\sigma^2_0$ &  & 14677.8 (5491.3, 25740.3) & 9139.5 (3044.4, 18879.9)\\
    $\sigma^2_1$ &  &  & 275.1 (102.1, 1164.2)\\
    $\phi_0$ &  & 11.6 (1.9, 29.2) & 18.7 (4.2, 29.4)\\
    $\phi_1$ &  &  & 0.1 (0.03, 3.4)\\
    $\tau^2$ & 22806.6 (16330.1, 32899.5) & 7105.5 (1999, 18985.2) & 5721.6 (1927.2, 15421.3)\\
    \bottomrule
  \end{tabular}
  \end{center}
\end{table}

Parameter estimates for candidate models are given in Table~\ref{tab:estimates}. As expected, the $\beta_1$ estimates indicate a strong positive relationship between 2012 mean canopy height and 2020 growing stock volume. The addition of spatial random effects decreased the non-spatial residual variance $\tau^2$. The reapportionment of $\tau^2$ to $\sigma^2_0$ and non-negligible $\sigma^2_1$ suggests a substantial portion of variance---not explained by the ALS predictor---had a spatial structure and the ALS predictor had space-varying impact.

The SVI model spatial decay parameter estimates suggest a fairly localized spatial structure. Given the exponential spatial correlation function and km map projection units, the distance $d_0$ at which the spatial correlation drops to 0.05 (an arbitrary, but commonly used, cutoff) is estimated by solving $0.05 = \exp(-\phi d_0)$ for $d_0$ providing $d_0=-\log(0.05)/\phi$. The distance $d_0$ is commonly referred to as the effective spatial range. Using the SVI model $\phi_0$ estimates, the corresponding spatial range is 0.26 (0.10, 1.58) km. 

Compared with the SVI model, the SVC model further reduced the non-spatial residual variance by taking into account the space-varying impact of $x_1$. Relative to the effective spatial range of the intercept process, the process on $x_1$ had a long spatial range, i.e.,  29.96 (0.88, 99.86) km. That is, the spatially varying slope coefficient for $x_1$ represented sub-region differences in the relationship between growing stock volume and mean canopy height that were probably caused by unmeasured species, genetic, or environmental factors. 

\begin{table}[!th]
  \caption{Candidate model fit and leave-one-out (LOO) cross-validation prediction diagnostics. The last three rows were calculated using prediction LOO on the observed data. The row labeled CI Cover is the percent of 95\% posterior predictive distribution credible intervals that cover the observed LOO value. Where appropriate, the ``best'' metric in the row is bolded. }\label{tab:fitPredict}
    \begin{center}
        \begin{tabular}{lccc}
        \toprule
        Model & Non-spatial & SVI & SVC\\
        \midrule
        DIC & 801.1 & 752.8 & \textbf{748.7}\\
        p$_D$ & 3.1 & 30.4 & 33.3\\
        L & -397.5 & -346.0 & -341.0\\
        \midrule
        WAIC$_1$ & 800.7 & 746.2 & \textbf{738.3}\\
        WAIC$_2$ & 801.1 & 763.8 & \textbf{756.9}\\
        p$_1$ & 2.8 & 23.8 & 22.9\\
        p$_2$ & 2.9 & 32.6 & 32.2\\
        LPPD & -397.6 & -349.3 & \textbf{-346.2}\\
        \midrule
        MSPE & 23378.4 & 23134.4 & \textbf{21491.6}\\
        CRPS & 88.2 & 87.0 & \textbf{82.9}\\
        CI cover & 98.4 & 96.8 & 98.4\\
    \bottomrule
    \end{tabular}
    \end{center}
\end{table}

Candidate model fit and LOO cross-validation prediction diagnostics are given in Table~\ref{tab:fitPredict}. Following from Section~\ref{sec:selection}, the lower values of DIC, WAIC$_1$, and WAIC$_2$ indicate addition of spatial random effects to the model intercept and regression slope coefficient improved fit to observed data. Similarly, LOO cross-validation MSPE and CRPS favored the SVC model over the Non-spatial and SVI model. All models achieve an approximate 95\% credible interval coverage.

\subsection{Blowdown prediction}

\begin{figure}[!ht]
\begin{center}
	\subfigure[]{\includegraphics[width=7.75cm,trim={0cm 0cm 0cm 0cm},clip]{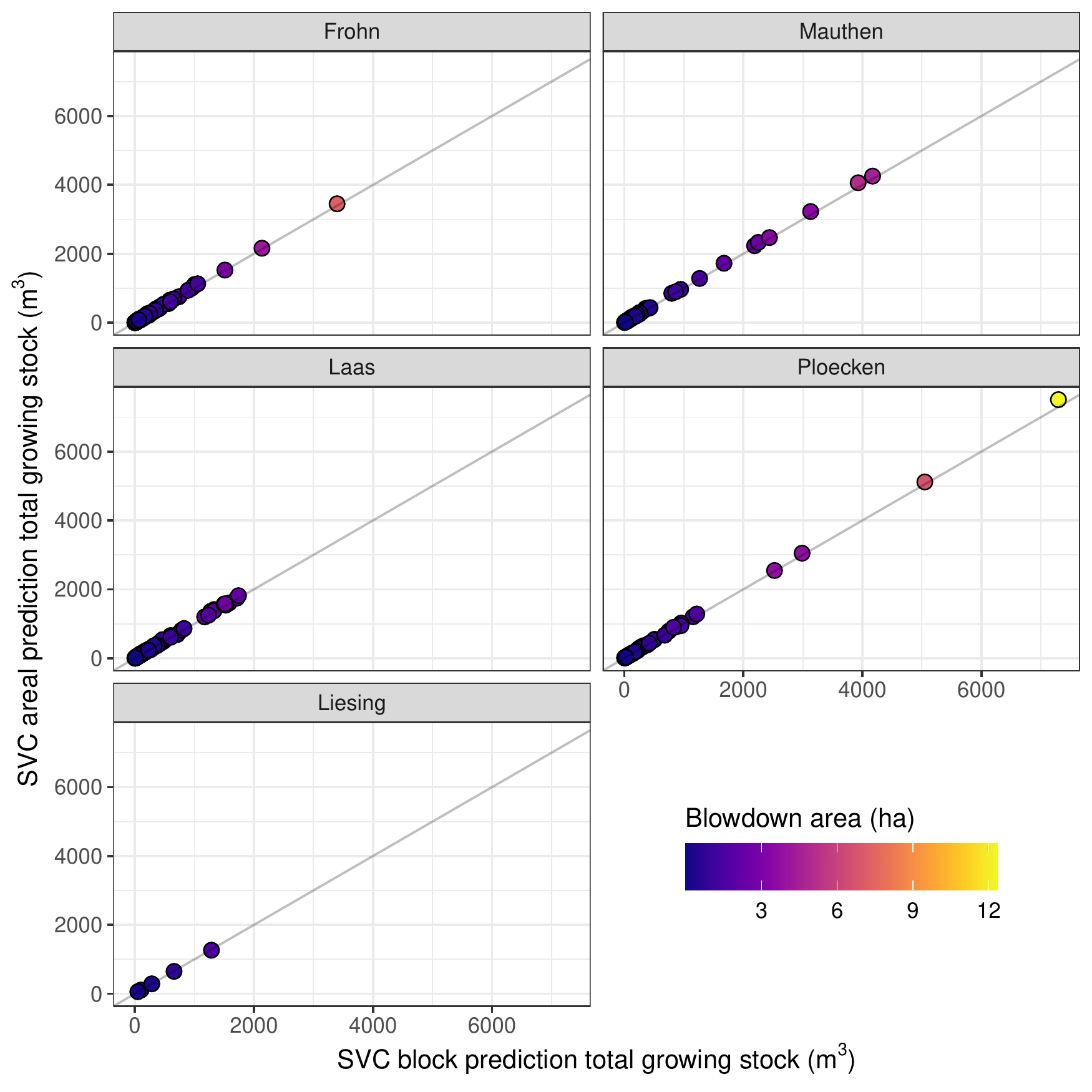}\label{fig:TotalPPDMean}}
	\subfigure[]{\includegraphics[width=7.75cm,trim={0cm 0cm 0cm 0cm},clip]{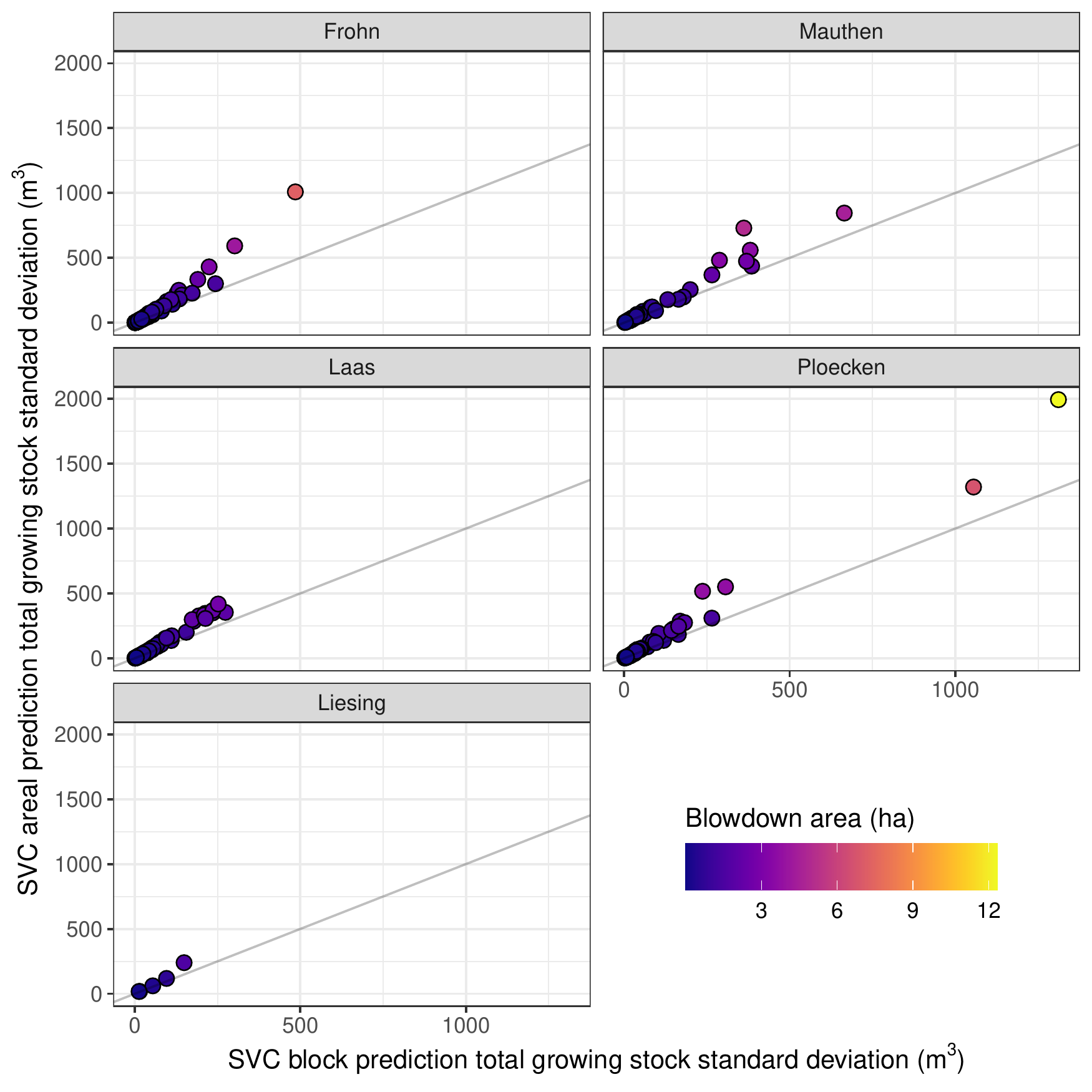}\label{fig:TotalPPDSD}}
	\caption{Summaries of each blowdown's total growing stock volume (m$^3$) posterior predictive distribution computed using areal and block prediction approach. Points represent blowdowns,  colored by area, and broken down by sub-region with a one-to-one line.} \label{fig:SVCTotalPPD}
	\end{center}
\end{figure}

\begin{figure}[!ht]
\begin{center}
	\includegraphics[width=7cm,trim={0cm 0cm 0cm 0cm},clip]{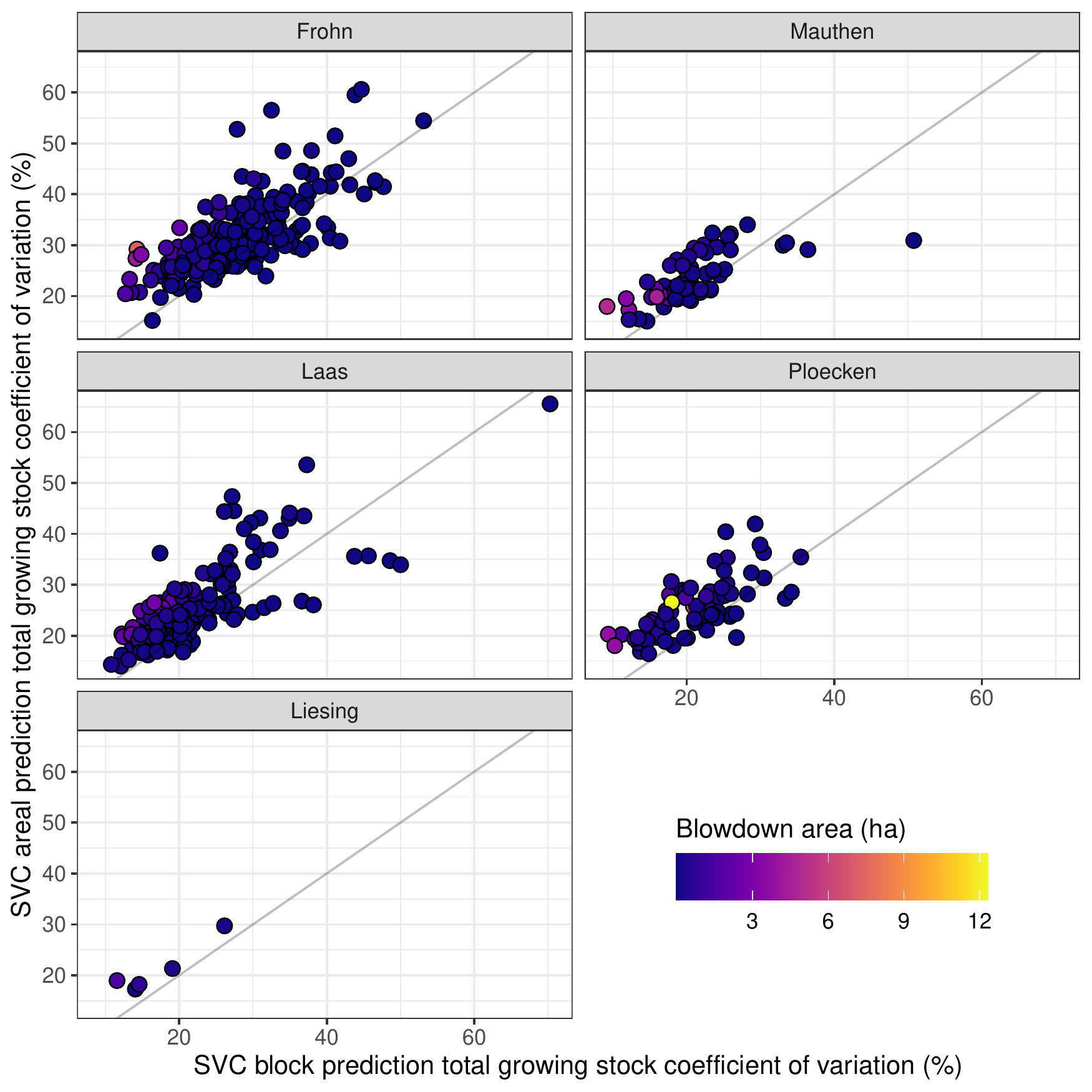}
	\caption{Summaries of each blowdown's total growing stock volume (m$^3$) posterior predictive distribution computed using areal and block prediction approach. Points represent blowdowns, colored by area, and broken down by sub-region with a one-to-one line.} \label{fig:TotalPPDCV}
	\end{center}
\end{figure}

The SVC model provided the ``best'' fit and LOO predictive performance (Table~\ref{tab:fitPredict}) and therefore served as the prediction model for the blowdowns. Following methods in Section~\ref{sec:blowdownPrediction}, areal and block growing stock volume PPD mean and standard deviation were computed for each blowdown, the results of which are plotted in Figure~\ref{fig:SVCTotalPPD}. Figure~\ref{fig:TotalPPDMean} shows negligible difference between areal and block PPD means. However, as shown in Figure~\ref{fig:TotalPPDSD}, compared with the block approach, the areal prediction resulted in a consistently larger PPD standard deviation. Additionally, supplemental Figure~\ref{fig:TotalPPDCV} shows the areal PPD coefficient of variation (CV) is generally larger than the block PPD CV.

Table~\ref{tab:predTotals} provides sub-region growing stock volume loss totals and corresponding 95\,\% confidence intervals. Although total blowdown area in Frohn was larger than other sub-regions, its per unit area growing stock loss 510.68\,m$^3$/ha (32,086.1\,m$^3$/62.83\,ha) was less than that in Laas (637.26\,m$^3$/ha), Liesing (818.21\,m$^3$/ha), Mauthen (784.76\,m$^3$/ha), and Ploecken (638.50\,m$^3$/ha). This disparity between Frohn and the other sub-regions is because the blowdowns in the Frohn sub-region were concentrated in relatively unproductive forests close to the alpine treeline zone.

\begin{table}[ht!]
  \caption{Growing stock volume loss by sub-region and study region posterior predictive distribution median and 95\% credible interval.}\label{tab:predTotals}
  \begin{center}
\begin{tabular}{lcc}
\toprule
  & Area (ha) & Volume (m$^3$)\\
\midrule
Frohn & 62.83 & 32086 (28290, 35921)\\
Laas & 54.95 & 35017 (31293, 38930)\\
Liesing & 2.90 & 2373 (1941, 2849)\\
Mauthen & 38.49 & 30205 (25676, 34758)\\
Ploecken & 53.15 & 33936 (28998, 39361)\\
\midrule
Total & 212.32 & 133775 (122935, 144308)\\
\bottomrule
\end{tabular}
\end{center}
\end{table}

\begin{figure}[!ht]
	\begin{center}
		\subfigure[PPD mean]{\includegraphics[width=5cm,trim={2.25cm 0cm 2.75cm 0cm},clip]{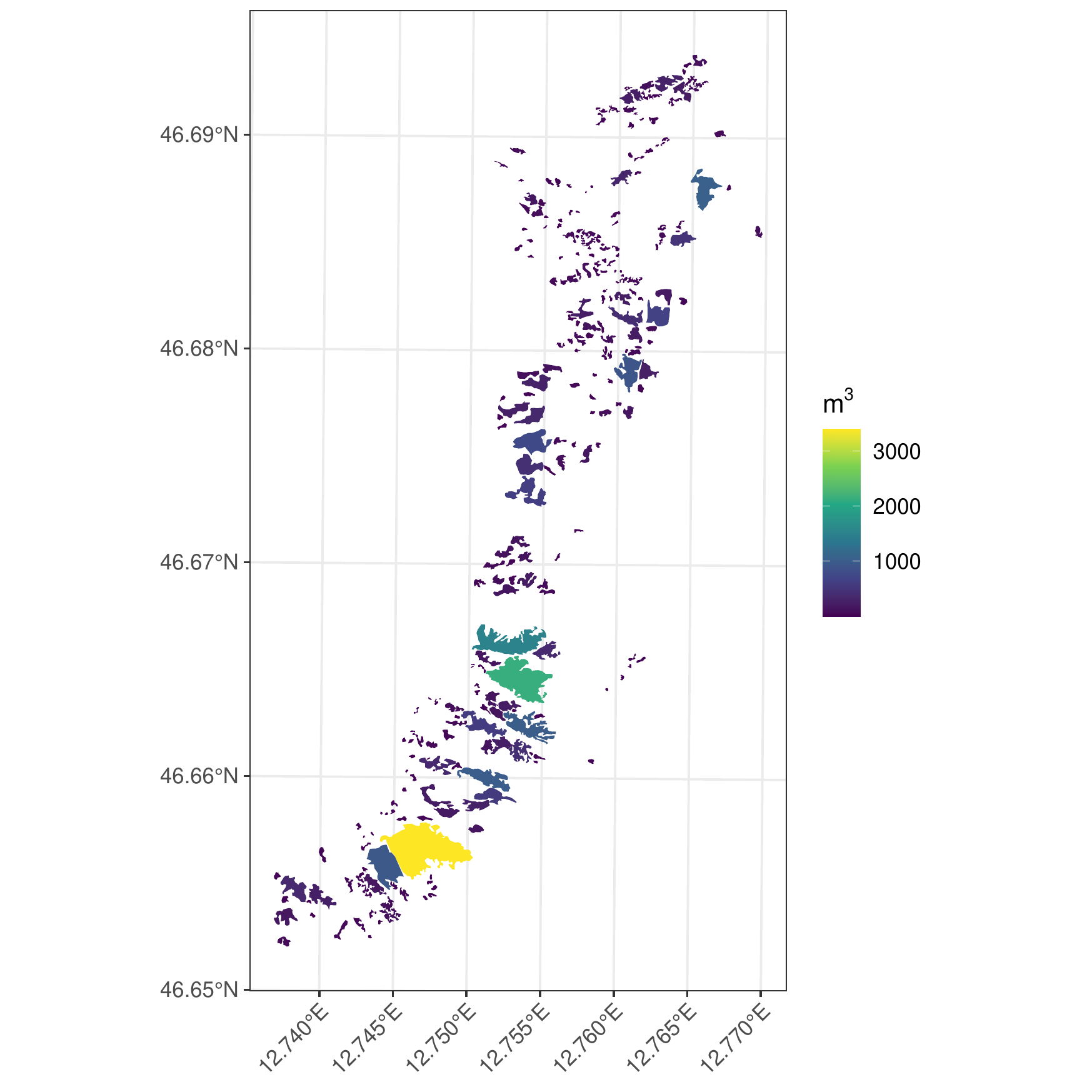}}
		\subfigure[PPD standard deviation]{\includegraphics[width=5cm,trim={2.25cm 0cm 2.75cm 0cm},clip]{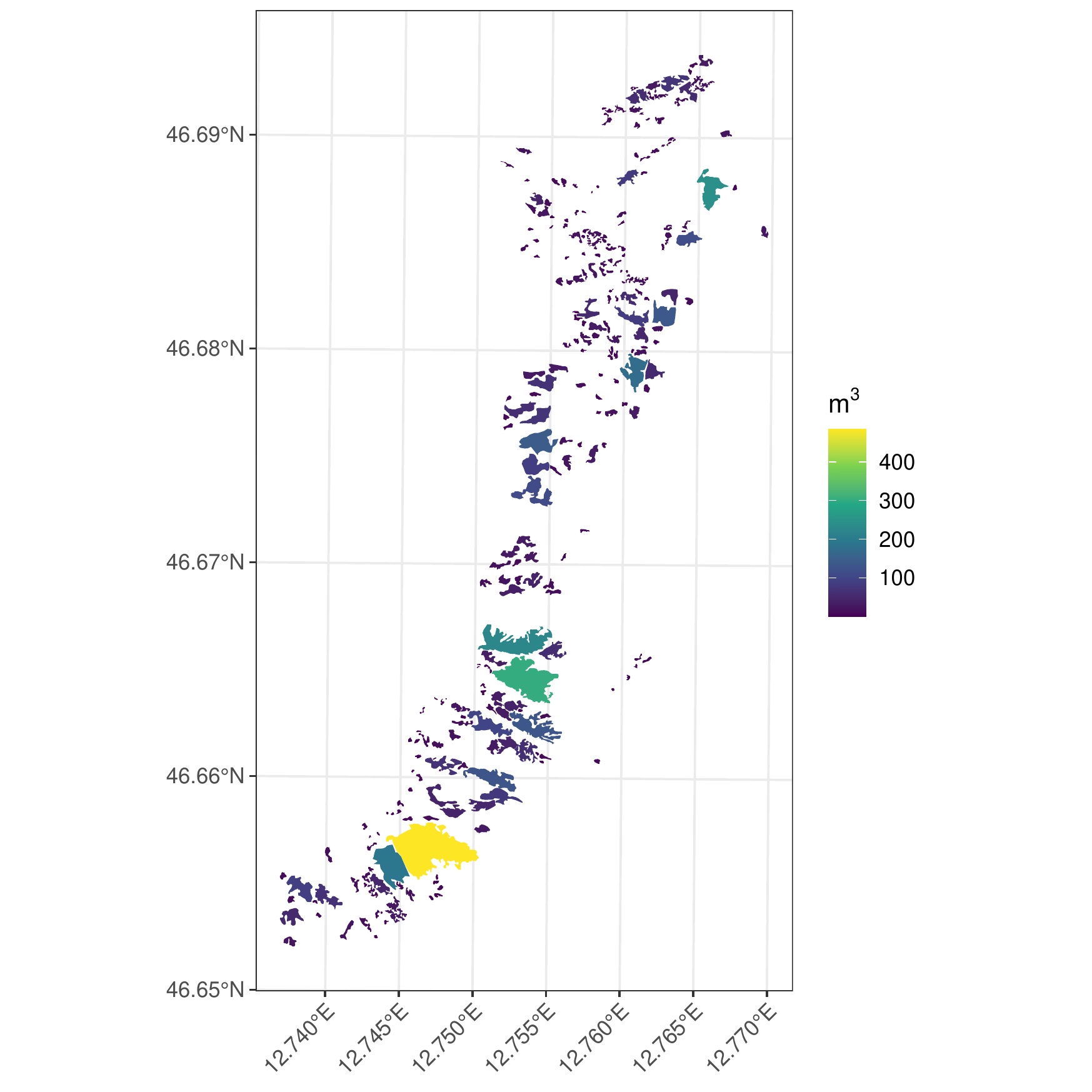}}
		\subfigure[PPD CV]{\includegraphics[width=5cm,trim={2.25cm 0cm 2.75cm 0cm},clip]{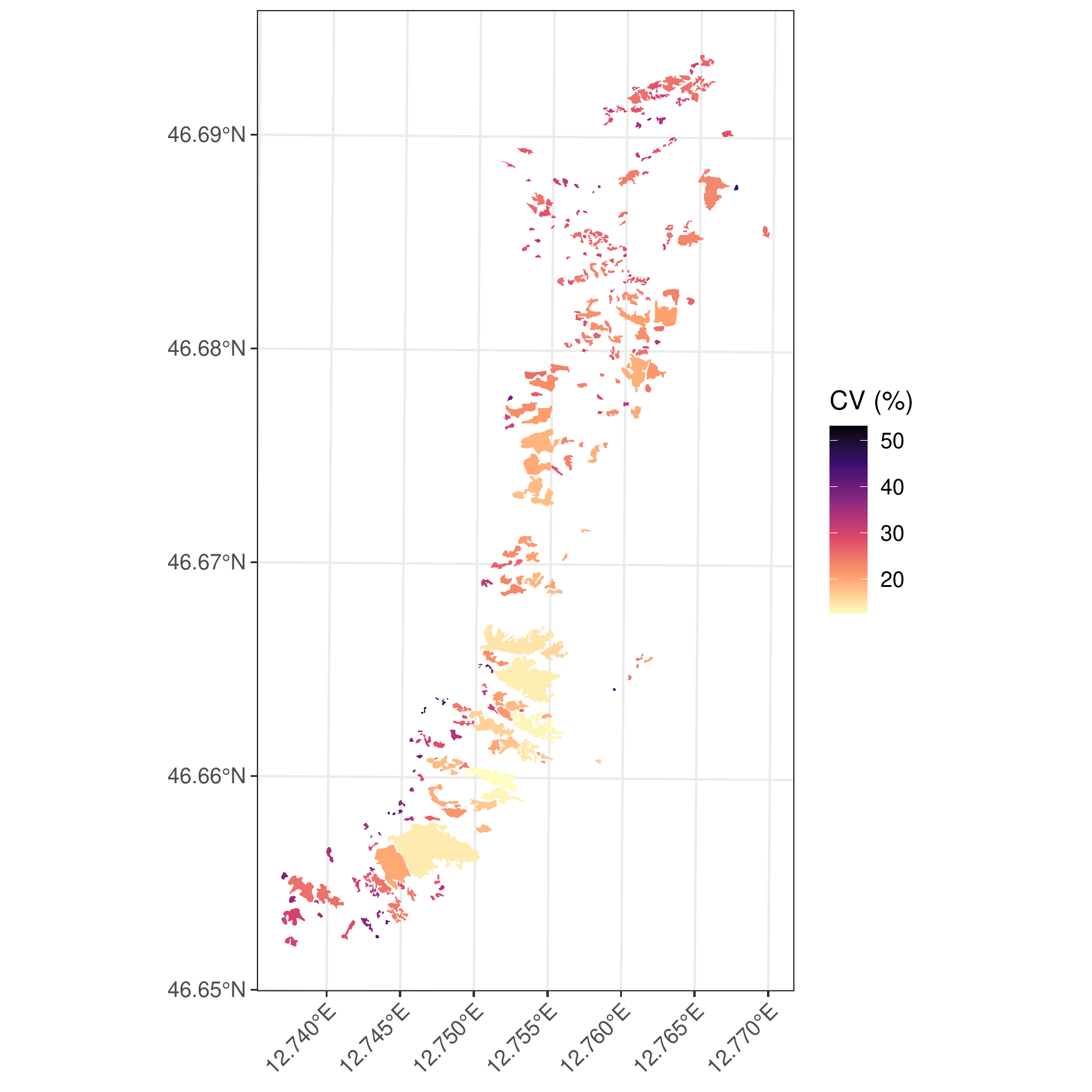}}
	\end{center}
	\caption{SVC block prediction approach posterior predictive distribution (PPD) mean, standard deviation, and coefficient of variation (CV) of total growing stock volume (m$^3$) for blowdowns in Frohn sub-region.} \label{fig:FrohnPred}
\end{figure}

Maps of PPD summaries per blowdown were prepared for each sub-region. For example, Figure~\ref{fig:FrohnPred} shows blowdowns in the Frohn sub-region colored by growing stock volume PPD estimate mean, standard deviation, and coefficient of variation. Similar maps for the other sub-regions are provided in the Supplemental Material.

\begin{figure}[!ht]
	\begin{center}
		\includegraphics[width=10cm,trim={0cm 0cm 0cm 0cm},clip]{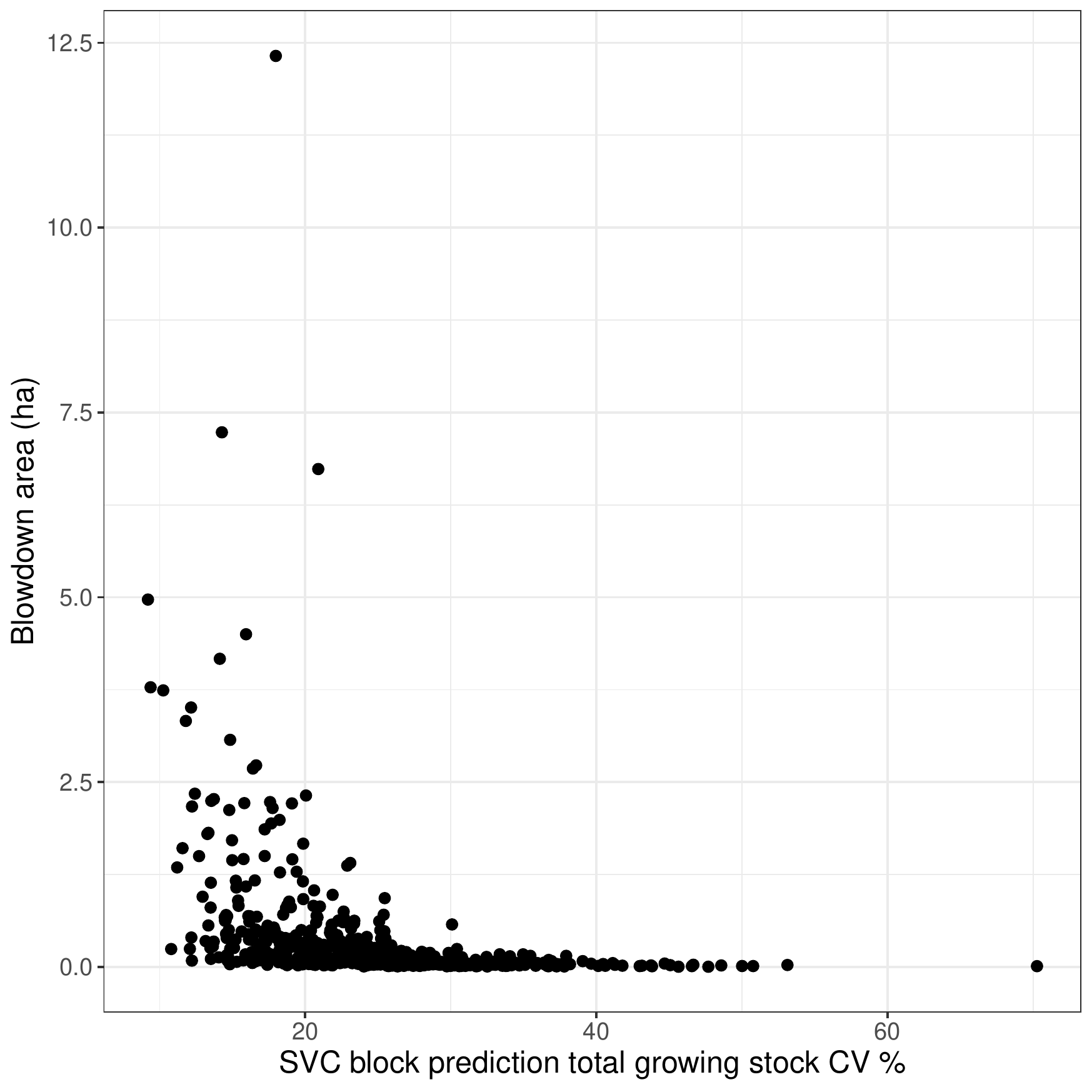}
	\end{center}
	\caption{SVC block prediction approach coefficient of variation (CV) posterior predictive distribution versus blowdowns area.} \label{fig:blkCV}
\end{figure}

Figure~\ref{fig:blkCV} shows the distribution of PPD CV by blowdown area. The average predicted growing stock volume CV over all blowdowns was 24.6\,\%, with individual CV predictions ranging from 9.2\,\% to 70.3\,\%. Approximately 90\,\% of blowdown predictions had CVs lower than 33.9\,\%. Relatively large prediction errors (i.e., large CVs) occurred for the smaller blowdown areas. Blowdowns for which the CV was less than 20\,\% accounted for 73.2\,\% of the total damage in the study region, and areas with a CV less than 25\,\% represented 93.2\,\% of the total damage.

\section{Discussion}\label{sec:discussion}

Among the set of ALS predictors, mean canopy height alone explained a substantial portion of the response variable's variance and yielded the lowest LOO cross-validation MSPE. This finding is similar to those in other related studies. For example, \cite{BreidenbachAstrup2012} selected mean canopy height to predict above-ground biomass using Norwegian national forest inventory data, and \cite{Magnussen2014} selected mean canopy height to predict growing stock density in a study that considered Swiss and Norwegian national forest inventory data. Further, \cite{Mauro2016} identified maximum vegetation height as the only significant predictor in small area estimation models used to improve quadratic mean diameter estimates in Central Spain.

Given the mean canopy height predictor, the SVC model showed consistently better fit and predictive performance compared to the Non-spatial and SVI models (Table~\ref{tab:fitPredict}), and was therefore selected as the model to generate areal and block prediction for the blowdowns.

In regression, and most other modeling contexts, it is assumed the predictor variables used to estimate model parameters are the same as those used in subsequent prediction. This assumption does not hold for the areal prediction approach described in Section~\ref{sec:blowdownPrediction}, because the ALS variables computed over the 62 0.126 ha sample plots could have a different distribution than the ALS variables computed over the variable area blowdowns. Although it is reasonable to expect the mean of a given ALS variable to be similar between the sample plot and blowdown distributions, the dispersion of the two distributions could be quite different. That is, a change-of-support was immanent between the sample plots (model data) and the prediction units' areal extent. This fact alone should precluded areal prediction application; however, our analysis results further underscore the approach's shortcomings. 

In the areal approach each blowdown was considered as a single prediction point, and its predicted volume per unit area was scaled by the blowdown's area to arrive at the blowdown's total volume. Such an approach might seem intuitive, but is problematic in the current setting. The issue is akin to how variance scales with forest inventory plot size. It is often the case that a forest variable's variance is inversely related to the area over which the variable is measured \citep{freese1962}. Consequently, a timber volume variable shows more variability when measured on smaller plots than on larger plots, especially when applied to structurally complex forests. This is because the larger plots average over local scale structural variability. When using a single prediction at the blowdown's centroid, the areal prediction variance does not scale with blowdown area. In contrast, the block approach more accurately scales prediction variance with blowdown area as reflected in Figure~\ref{fig:TotalPPDSD}. This is because its PPD is the result of an average over possibly multiple spatially correlated prediction units (grid cells) within the blowdown's boundary---the number of prediction units contributing to this average scales with blowdown area. 

Finally, the PPDs rely on the spatial correlation between observed and prediction locations, e.g., using Equation~\ref{matern}, to appropriately weight the contribution of observed data for prediction. For the areal approach, this correlation is computed between observed locations and a blowdown's polygon centroid, which is the average location relative to the polygon's vertices. If the polygon's shape is highly irregular, with a large boundary length to area ratio, then its centroid might not represent well the distance between observed locations and the polygon's extent. As an extreme example, the centroid of a polygon in the shape of a letter ``C'' or ``L'' will fall outside the polygon boundary. Again, such issues are circumvented using the block prediction approach.

A particularly useful quality of the Bayesian inferential paradigm is PPDs can be generated for any arbitrary set of prediction units from a single grid cell to sets of blowdowns that comprise a sub-region or entire study region. Inference proceeds from desired joint PPD samples. These PPD samples can be summarized using measures of central tendency, dispersion, intervals, and transformations, with results presented as maps or tables, e.g., Figure~\ref{fig:FrohnPred} and Table~\ref{tab:predTotals}. 

\section{Conclusions}\label{sec:conclusion}

The study goal to estimate growing stock timber volume loss due to storm Adrian in the Austrian upper Gail valley in Carinthia, was met using ALS and TLS measurements coupled through a flexible spatial regression model cast within a Bayesian inferential framework. Limited data availability and its configuration in space and time presented several inferential constraints on how statistically robust estimates could be pursued. Our proposed regression model was designed to leverage information from a small set of plot samples and aerial ALS as well as spatial autocorrelation among forest measurements and nonstationary relationships between response and predictor variables. 

Three candidate models of varying complexity were assessed using model fit and out-of-sample prediction. Performance metrics supported the SVC model which was then used to make areal and block predictions over the blowdowns. Our results showed that in contrast to the areal approach, the block approach mitigates issues with change-of-support by matching the prediction unit to the sample plot extent. Using this finer spatial scale prediction unit and a joint prediction algorithm that acknowledges spatial correlation among prediction units, the total growing stock volume PPD captures the correlation among prediction units within a given blowdown and scale appropriately with blowdown area. The block prediction approach facilitated statistically sound inference at various spatial scales, i.e., blowdown, sub-region, and region levels. 

The proposed methodology and annotated code that yields fully reproducible results can be adapted to deliver damage assessment for other forest disturbance events in future periods and different geographic regions. While additional sample plot data would improve estimates in our current study, we were able to demonstrate a fairly high level of accuracy and precision is achievable using a limited sample size. This small sample, i.e., $n$=62 plots, was collected using a TLS which, compared with traditional individual tree measurements, allowed for time and effort efficient data collection. Based on our experience from this and other efforts, a field crew can collect $\sim$20-25 plots per day, even in difficult alpine terrain. 

Our study illustrated a methodology to efficiently deliver information required for strategic salvage harvesting following storm and other disturbances. Future work focuses on augmenting the SVC model to incorporate large spatial datasets using Nearest Neighbor Gaussian Processes \citep{Finley2019}, automate remote sensing predictor variable selection \citep{Franco-Villoria2019}, and accommodate high-dimensional \citep{Finley2017, Taylor-Rodriguez2019} and distributional regression \citep{Umlauf2018, Stasinopoulos2018} response vectors to predict forest disturbance induced change in species composition and diameter/size distributions. 

\section{Acknowledgments}

This study was supported by the project Digi4+ and was financed by the Austrian Federal Ministry of Agriculture, Regions and Tourism under project number 101470. Finley was supported by the United States National Science Foundation DMS-1916395 and by NASA Carbon Monitoring System grants. The authors appreciate the support during the fieldwork that was given by the forest owners, Clemens Wassermann, G{\"u}nter Kronawetter and the team of the Carinthian Forest Service.

\bibliography{NGKERSF21.bib}

\newpage
\beginsupplement
\section*{Supplemental material}

\begin{figure}[!ht]
	\begin{center}
		\subfigure[Posterior mean]{\includegraphics[width=7cm,trim={0cm 2cm 0cm 2cm},clip]{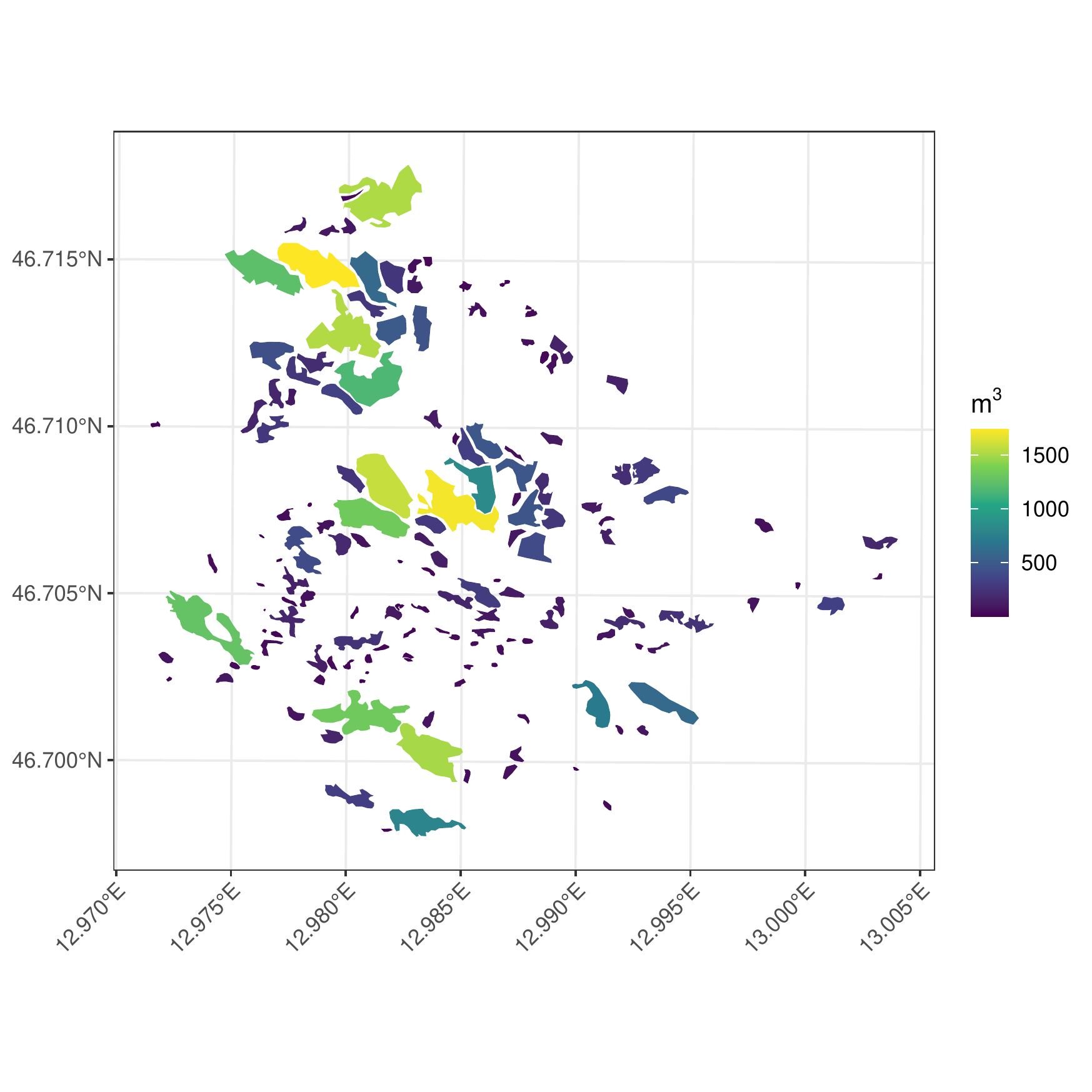}}
		\subfigure[Posterior standard deviation]{\includegraphics[width=7cm,trim={0cm 2cm 0cm 2cm},clip]{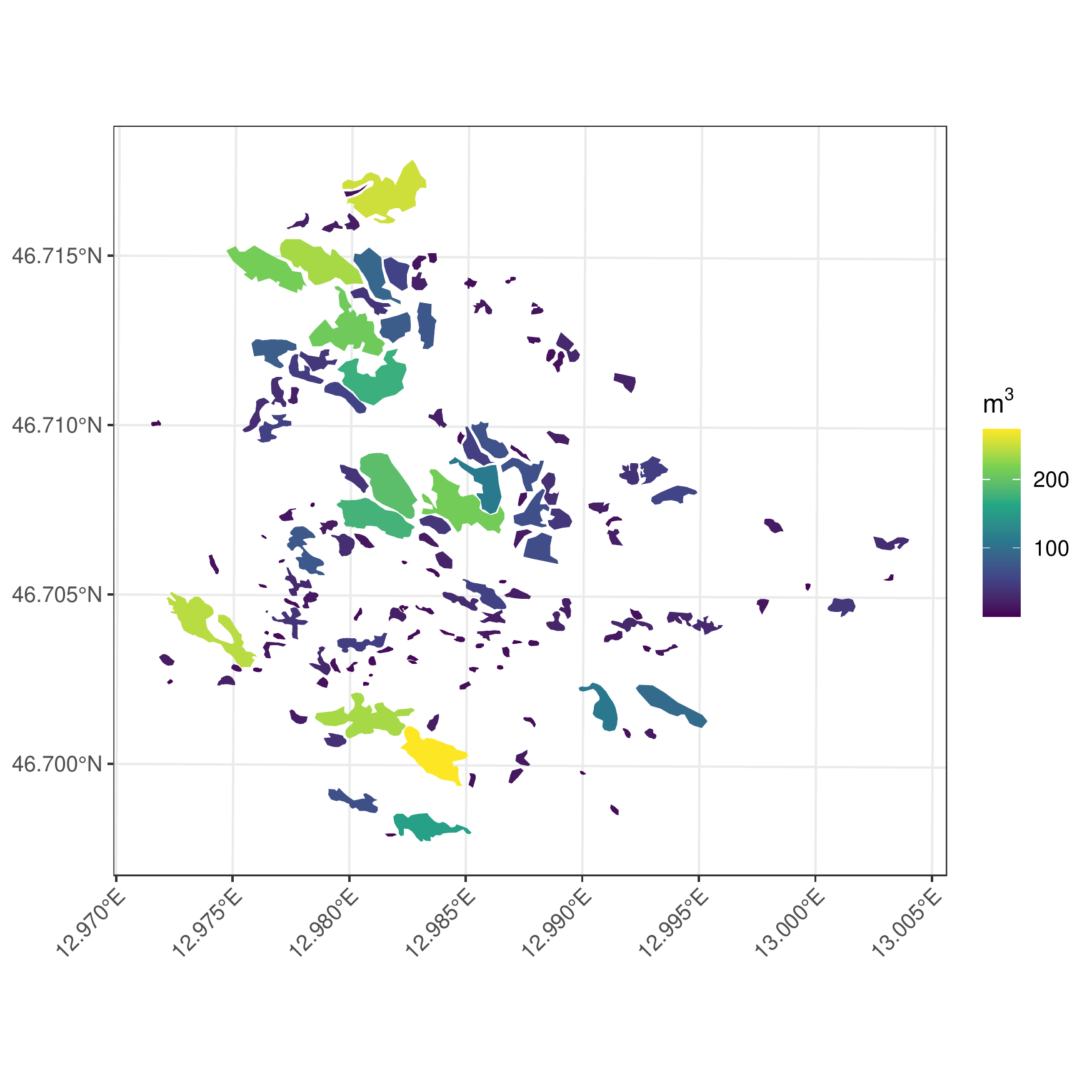}}\\
		\subfigure[Posterior coefficient of variation]{\includegraphics[width=7cm,trim={0cm 2cm 0cm 2cm},clip]{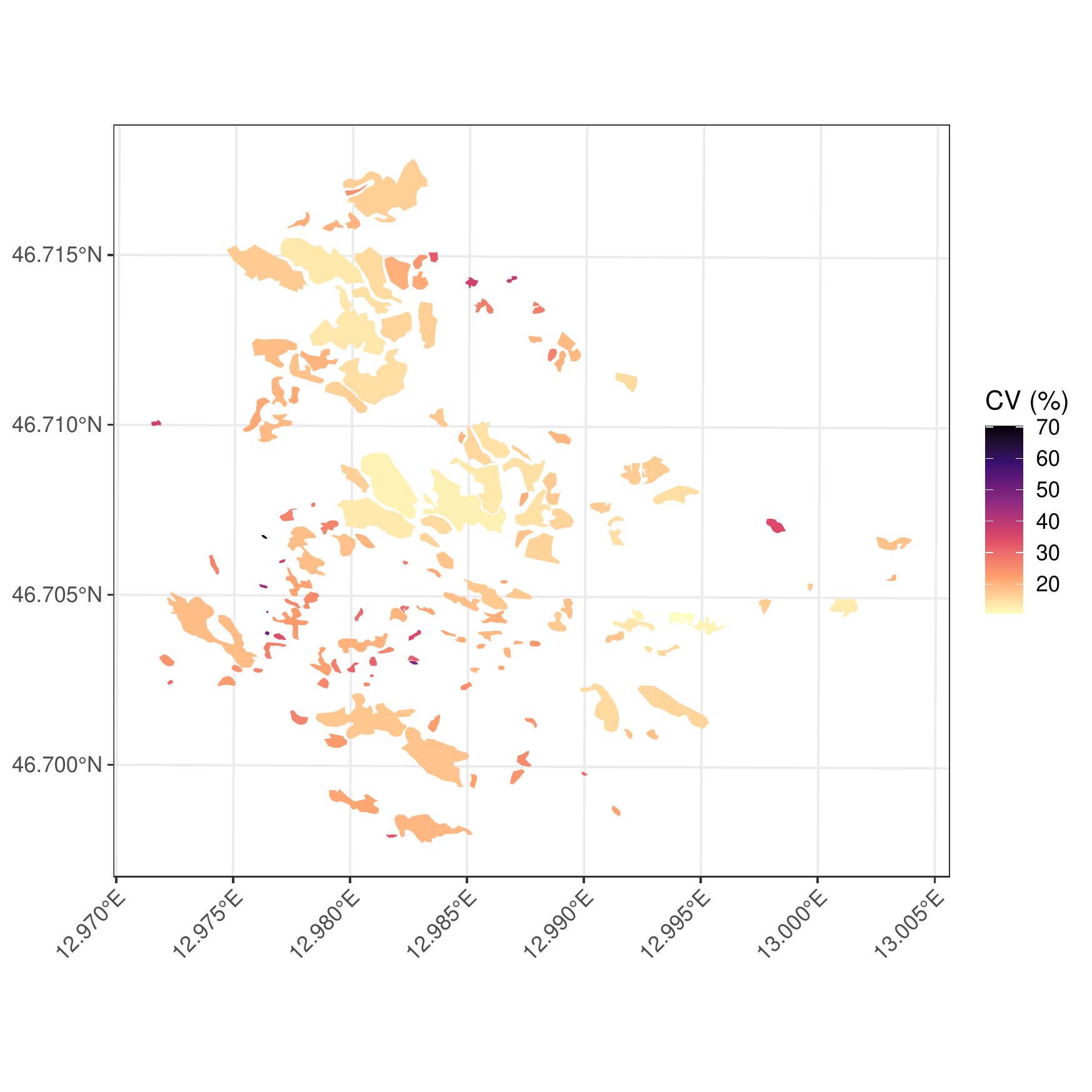}}
	\end{center}
	\caption{Posterior predictive distribution summaries for the blowdowns in Laas sub-region.} \label{fig:Laas_pred}
\end{figure}

\begin{figure}[!ht]
	\begin{center}
		\subfigure[Posterior mean]{\includegraphics[width=7cm,trim={0cm 0cm 0cm 0cm},clip]{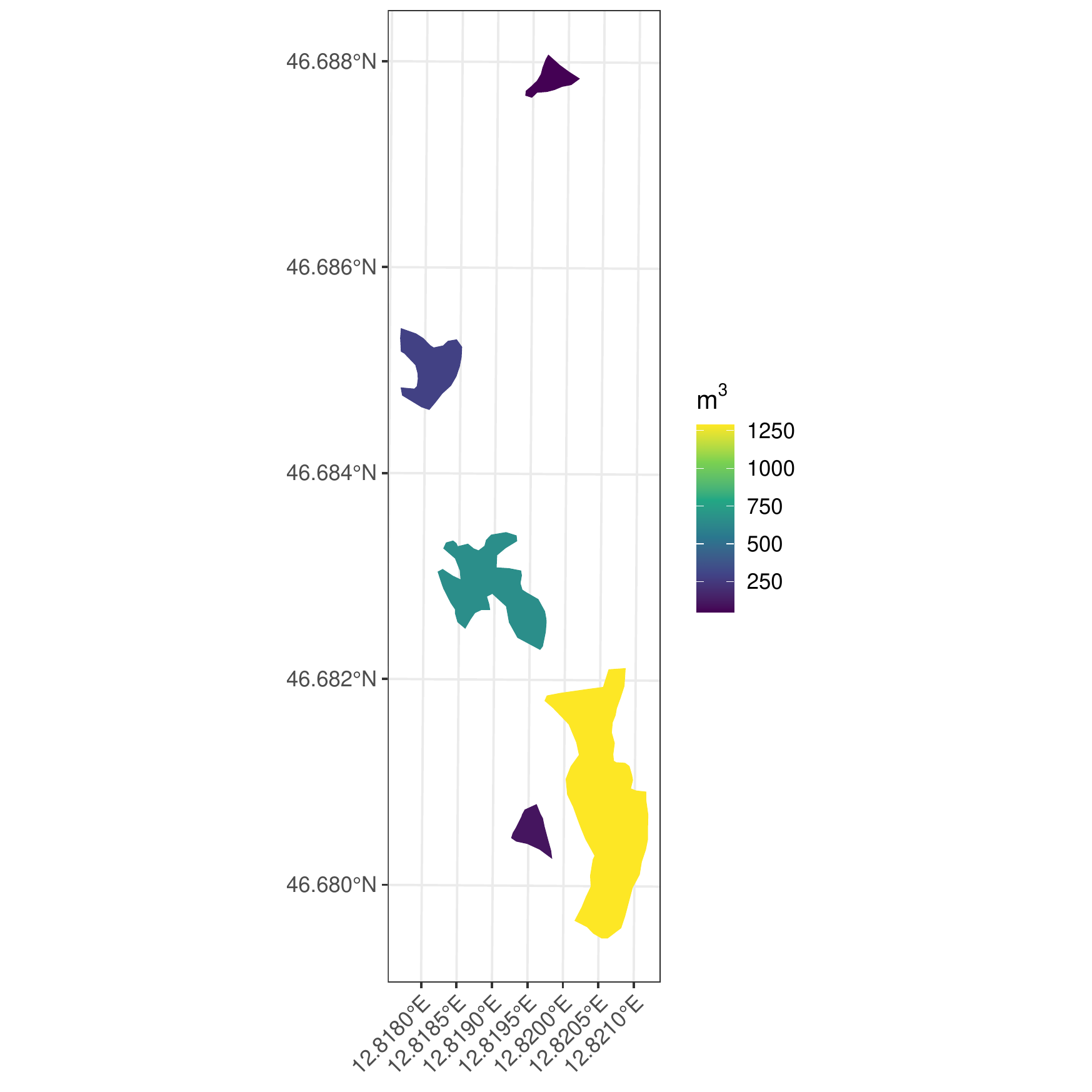}}
		\subfigure[Posterior standard deviation]{\includegraphics[width=7cm,trim={0cm 0cm 0cm 0cm},clip]{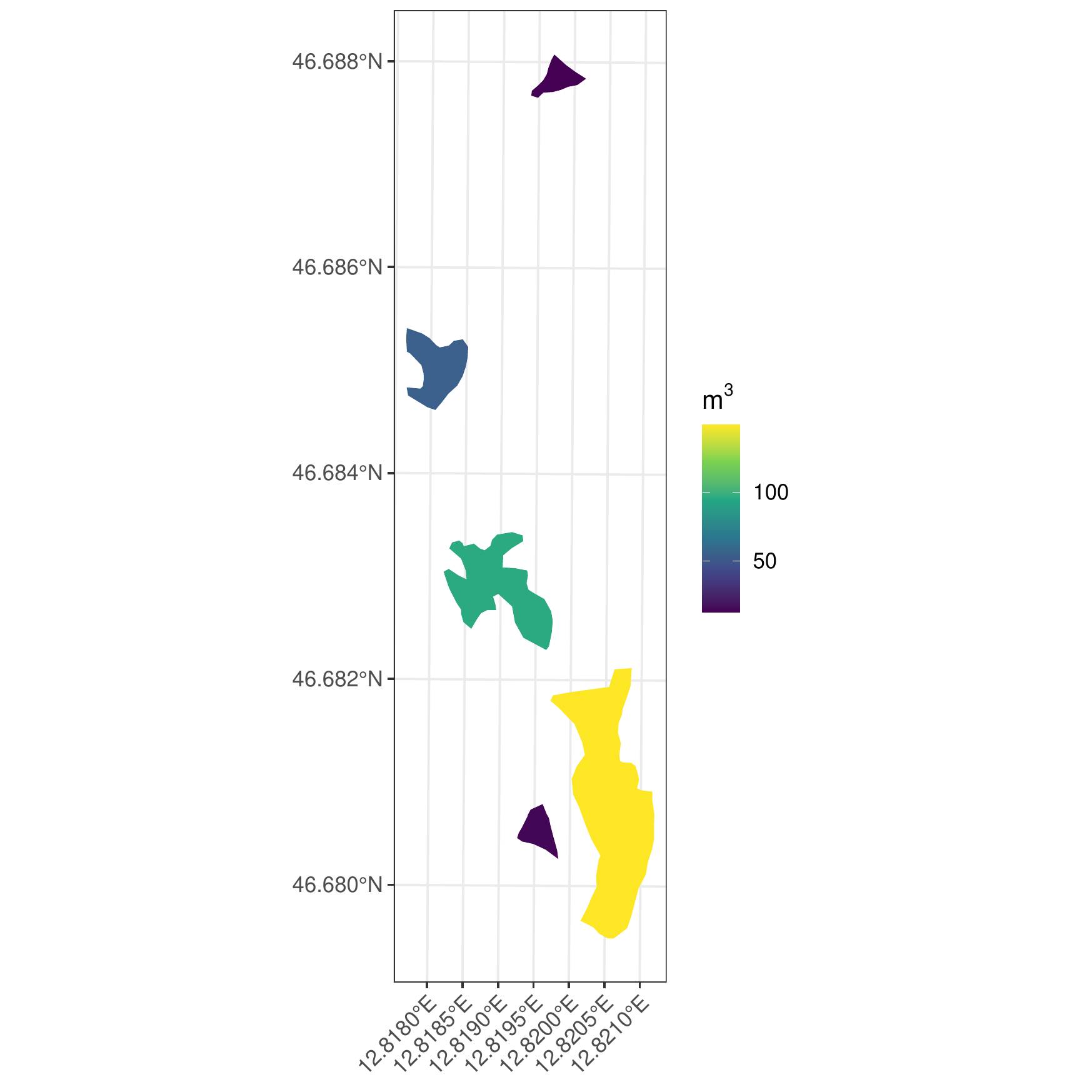}}\\
		\subfigure[Posterior coefficient of variation]{\includegraphics[width=7cm,trim={0cm 0cm 0cm 0cm},clip]{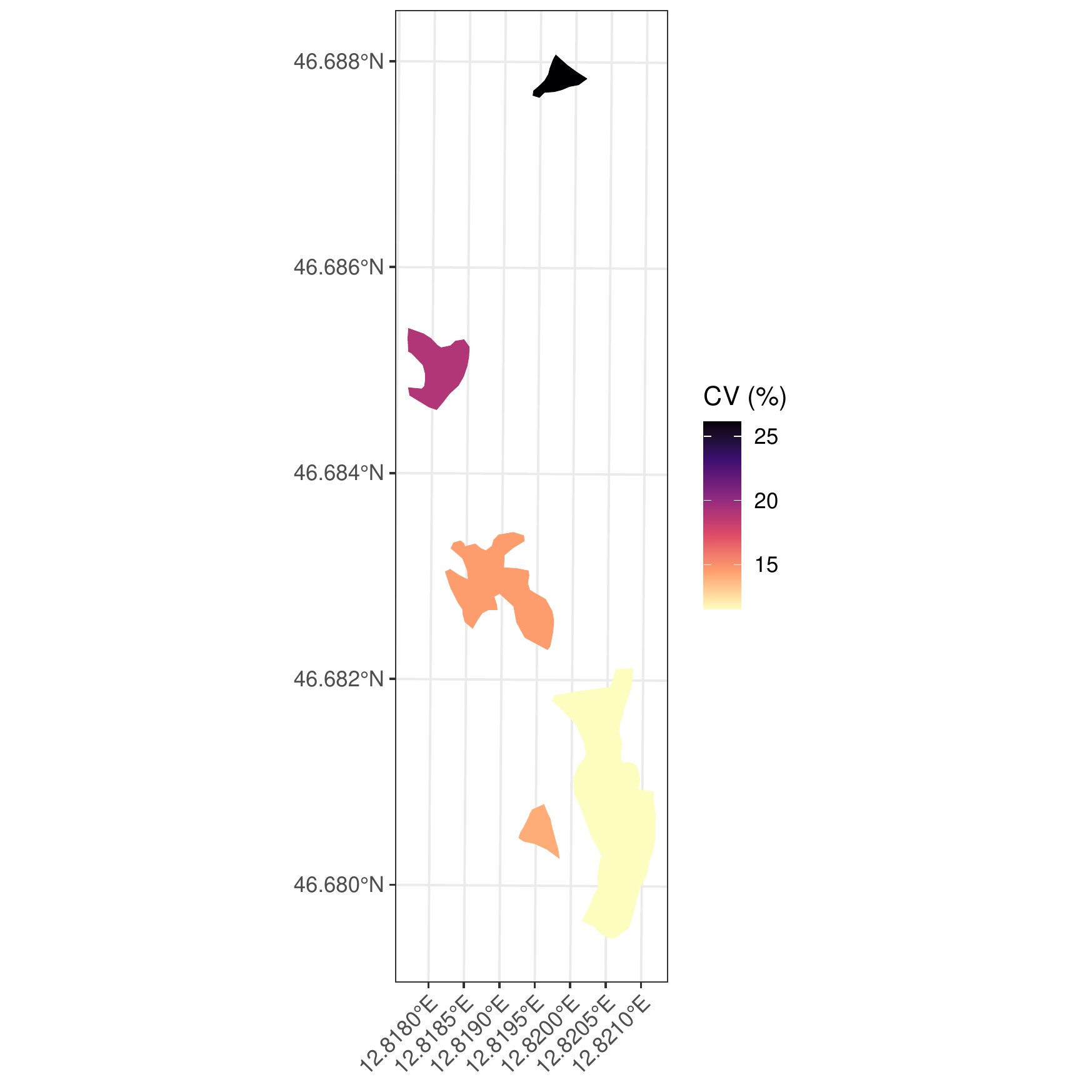}}
	\end{center}
	\caption{Posterior predictive distribution summaries for the blowdowns in Liesing sub-region.} \label{fig:LiesingPred}
\end{figure}

\begin{figure}[!ht]
	\begin{center}
		\subfigure[Posterior mean]{\includegraphics[width=7cm,trim={0cm 0cm 0cm 0cm},clip]{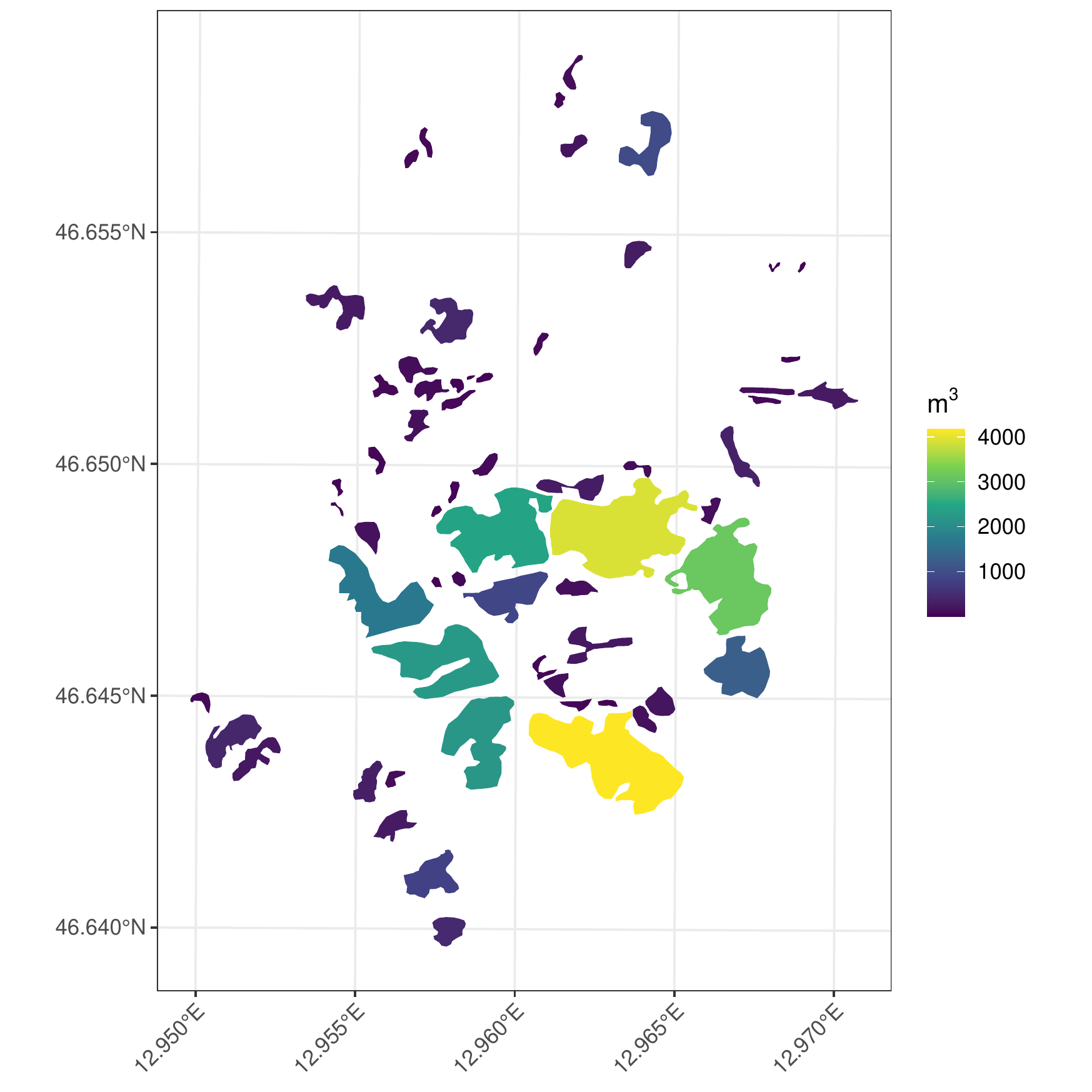}}
		\subfigure[Posterior standard deviation]{\includegraphics[width=7cm,trim={0cm 0cm 0cm 0cm},clip]{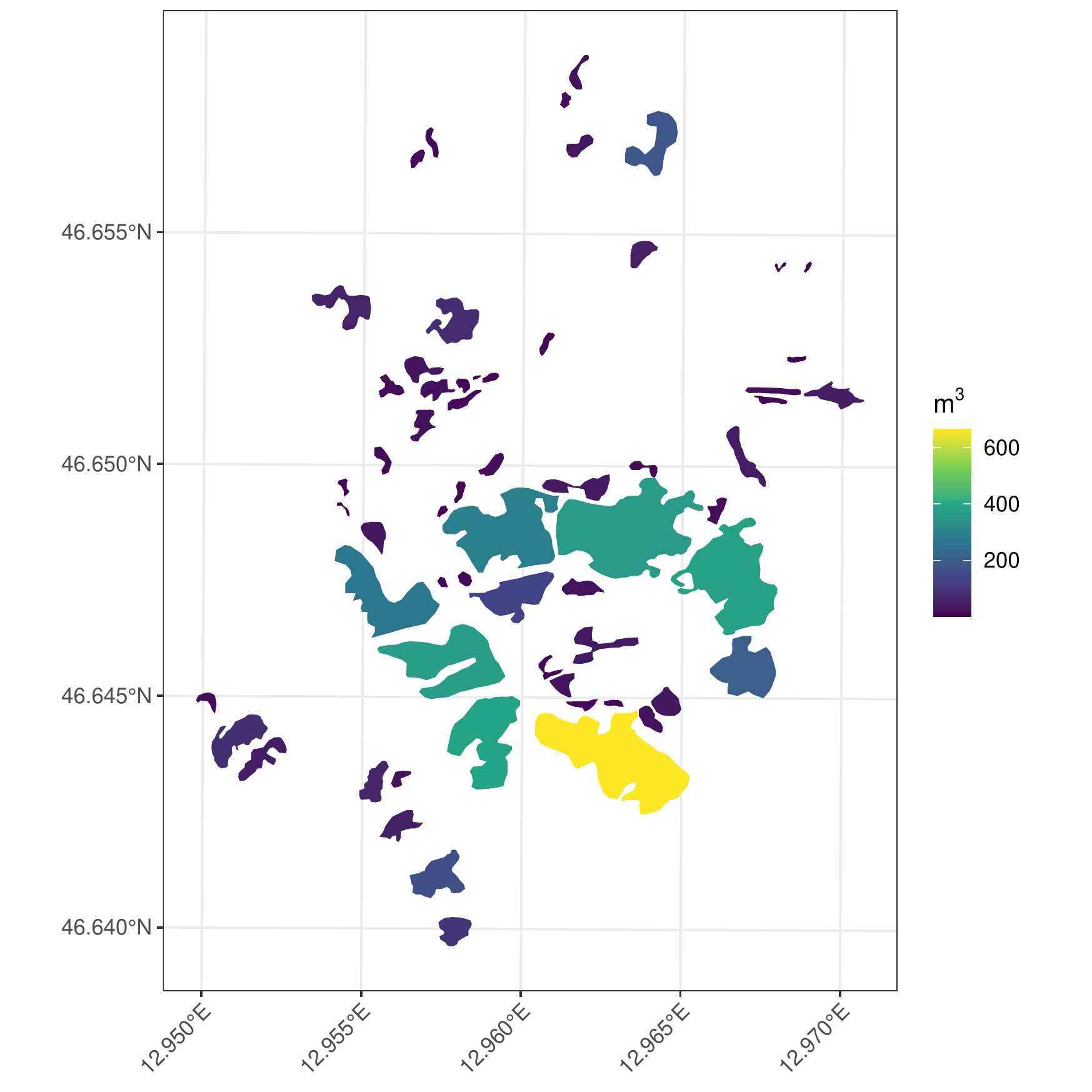}}\\
		\subfigure[Posterior coefficient of variation]{\includegraphics[width=7cm,trim={0cm 0cm 0cm 0cm},clip]{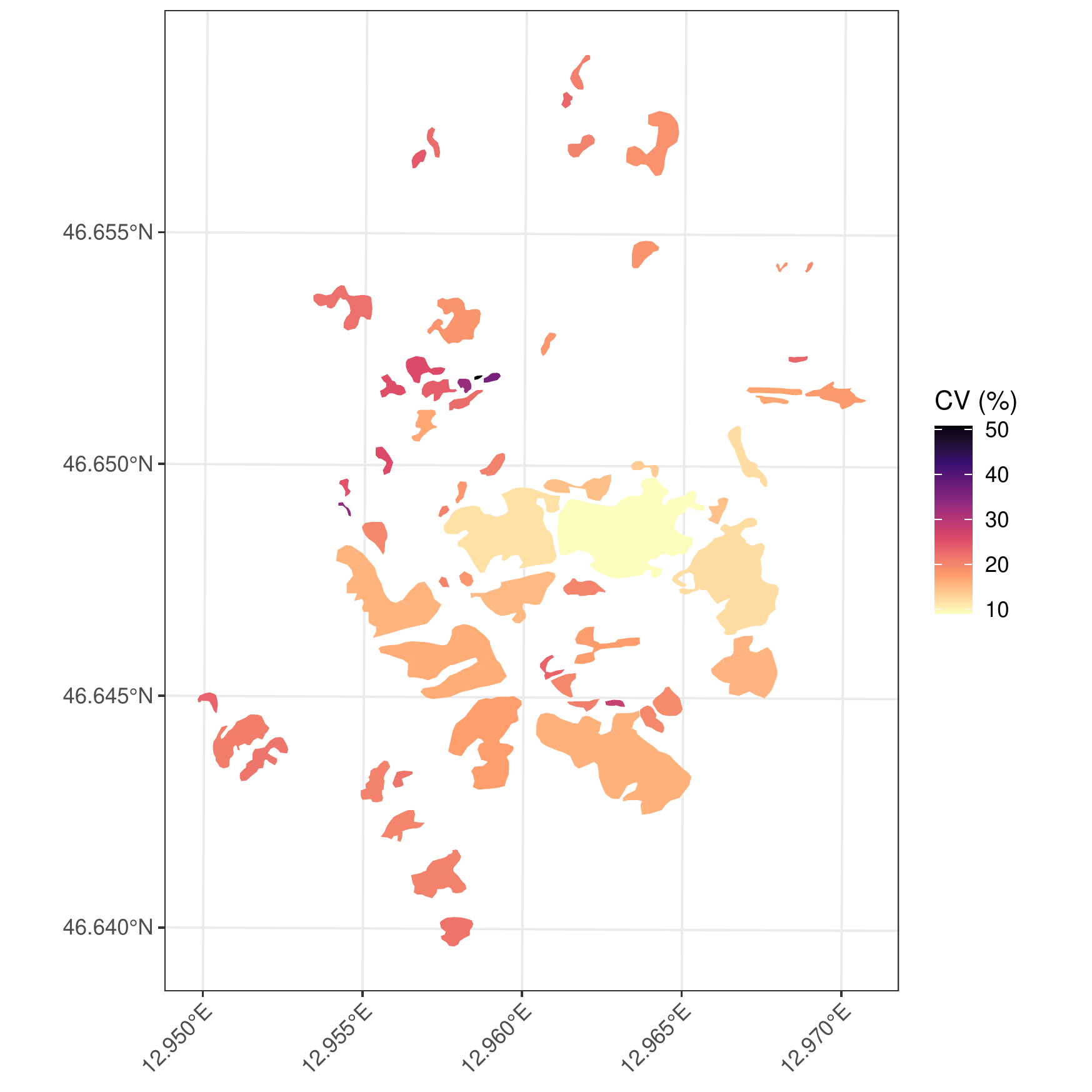}}
	\end{center}
	\caption{Posterior predictive distribution summaries for the blowdowns in Mauthen sub-region.} \label{fig:MauthernPred}
\end{figure}

\begin{figure}[!ht]
	\begin{center}
		\subfigure[Posterior mean]{\includegraphics[width=7cm,trim={0cm 4cm 0cm 4cm},clip]{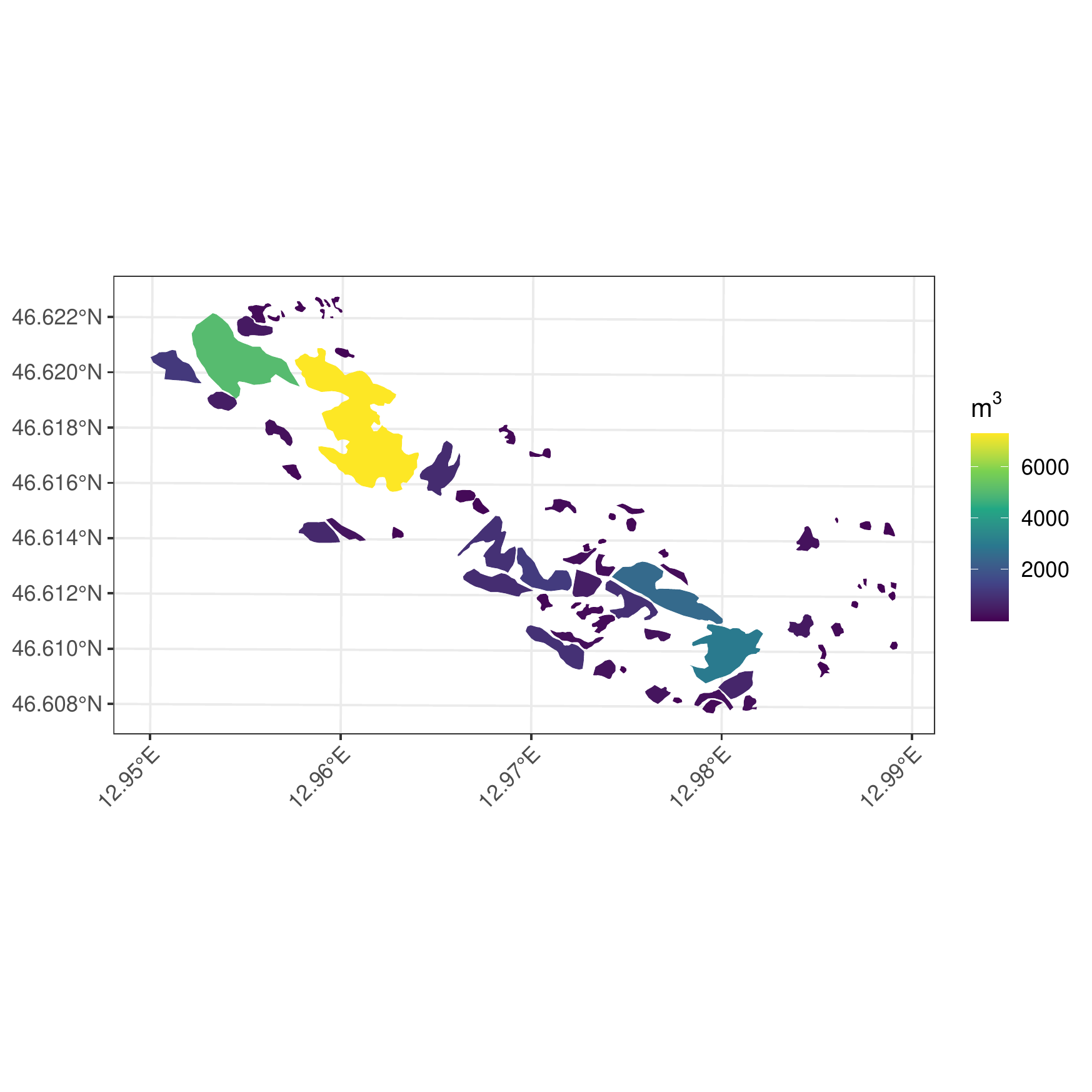}}
		\subfigure[Posterior standard deviation]{\includegraphics[width=7cm,trim={0cm 4cm 0cm 4cm},clip]{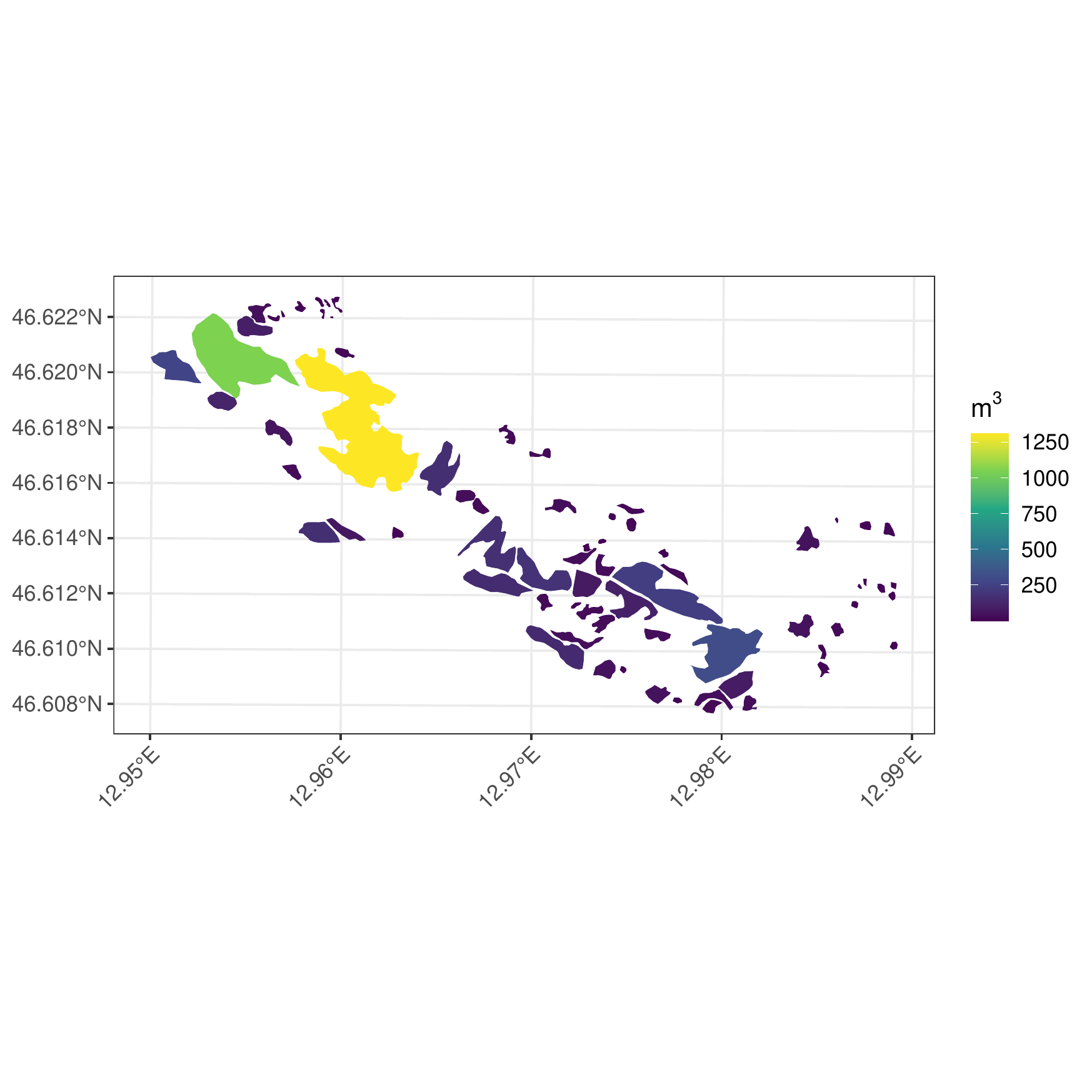}}\\
		\subfigure[Posterior coefficient of variation]{\includegraphics[width=7cm,trim={0cm 4cm 0cm 4cm},clip]{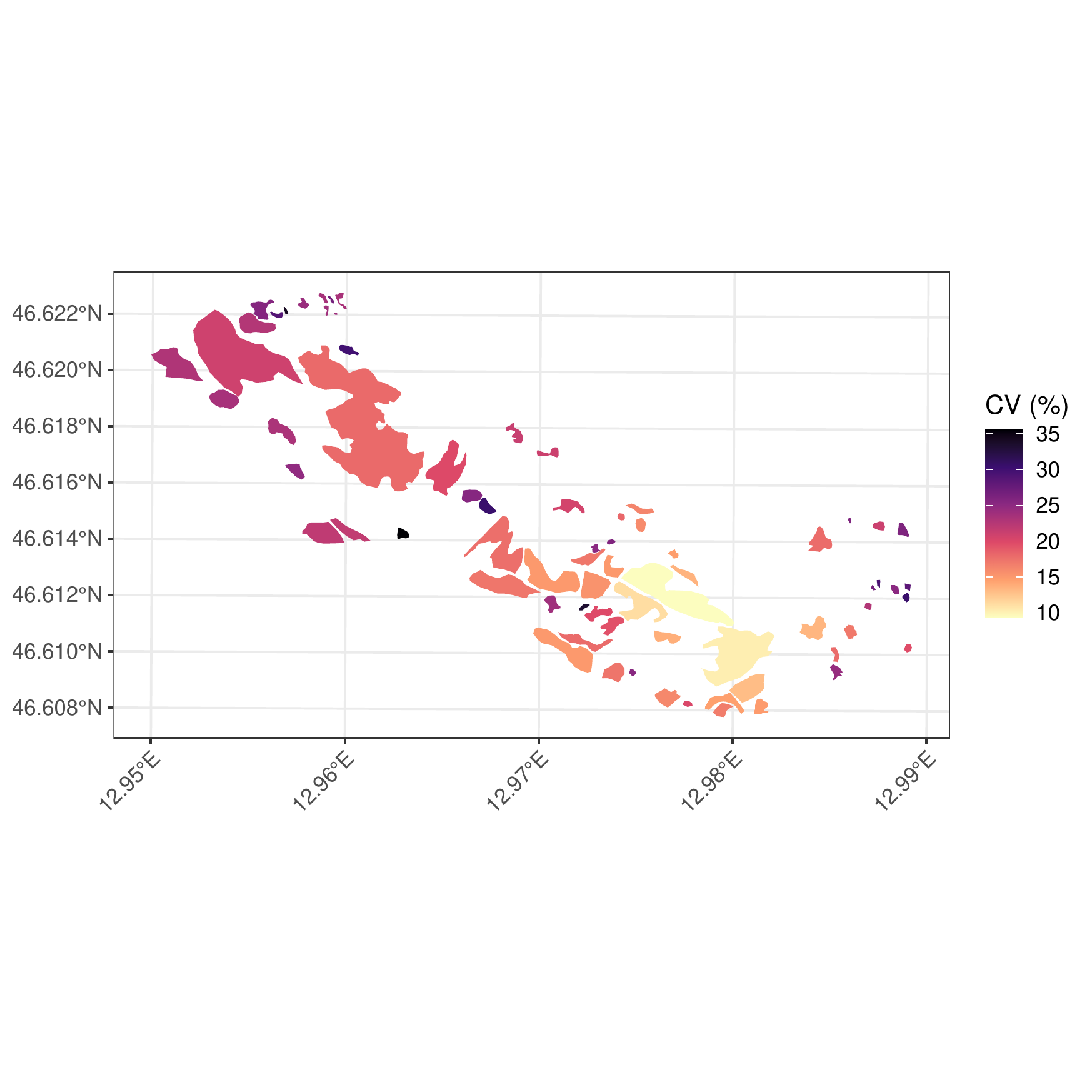}}
	\end{center}
	\caption{Posterior predictive distribution summaries for the blowdowns in Pl\"{o}cken sub-region.} \label{fig:PloeckenPred}
\end{figure}

\end{document}